\documentclass{ws-ijmpd2}
\usepackage[super,compress]{cite}

\usepackage{moresize}
\usepackage{enumerate}
\usepackage[caption=false]{subfig}
\usepackage{multirow}
\usepackage{mathtools}
\usepackage{stackengine}

\newcommand{\be}{\begin{equation}}
\newcommand{\ee}{\end{equation}}
\newcommand{\bea}{\begin{eqnarray}}
\newcommand{\eea}{\end{eqnarray}}
\newcommand{\bse}{\begin{subequations}}
\newcommand{\ese}{\end{subequations}}
\newcommand{\bce}{\begin{center}}
\newcommand{\ece}{\end{center}}
\newcommand{\bfg}{\begin{figure}}
\newcommand{\efg}{\end{figure}}
\newcommand{\bed}{\begin{description}}
\newcommand{\eed}{\end{description}}
\newcommand{\ben}{\begin{enumerate}}
\newcommand{\een}{\end{enumerate}}

\newcommand{\bit}{\begin{itemlist}}
\newcommand{\eit}{\end{itemlist}}

\newcommand{\nn}{\nonumber}

\newcommand{\la}{\label}
\newcommand{\pa}{\partial}
\newcommand{\fr}{\frac}
\newcommand{\sq}{\sqrt}
\newcommand{\no}{\noindent}

\newcommand{\lra}{\longrightarrow}

\def\a  {\alpha}
\def\b  {\beta}
\def\c  {\gamma}
\def\C  {\Gamma}
\def\d  {\delta}

\def\e  {\epsilon}

\def\f  {\phi}

\def\k  {\kappa}
\def\l  {\lambda}
\def\L  {\Lambda}
\def\m  {\mu}
\def\n  {\nu}

\def\O  {\Omega}

\def\r  {\rho}

\def\s  {\sigma}
\def\t  {\tau}
\def\vep {\varepsilon}
\def\vph {\varphi}

\def\le {\left}
\def\ri {\right}
\newcommand{\cA}{\mathcal A}

\newcommand{\cF}{\mathcal F}
\newcommand{\cG}{\mathcal G}

\newcommand{\cL}{\mathcal L}
\newcommand{\cM}{\mathcal M}
\newcommand{\cN}{\mathcal N}

\newcommand{\cQ}{\mathcal Q}
\newcommand{\cR}{\mathcal R}
\newcommand{\cS}{\mathcal S}
\newcommand{\cT}{\mathcal T}

\newcommand{\fM}{\mathfrak M}

\newcommand{\fw}{\mathfrak w}
\newcommand{\nab}{\nabla\!}
\newcommand{\nt}{\widetilde{\nabla}\!}

\newcommand{\Rt}{\widetilde{\cR}}
\newcommand{\Ct}{\widetilde{\C}}

\newcommand{\hg}{\widehat{g}}

\newcommand{\hR}{\widehat{\cR}}
\newcommand{\hS}{\widehat{\cS}}
\newcommand{\hLm}{\widehat{\cL}^{(m)}}

\newcommand{\keff}{\kappa_{\text{\scriptsize eff}}}
\newcommand{\Geff}{G_{\text{\scriptsize eff}}}

\newcommand{\rmt}{\r^{(m)}}

\newcommand{\rtt}{\r^{(\t)}}
\newcommand{\ptt}{p^{(\t)}}

\newcommand{\rx}{\r_{\!_X}}
\newcommand{\px}{p_{\!_X}}

\newcommand{\rJ}{\r_{\!_J}}
\newcommand{\pJ}{p_{\!_J}}

\newcommand{\Om}{\O^{(m)}}
\newcommand{\Otau}{\O^{(\t)}}

\newcommand{\OX}{\O_{\text{\ssmall {\it X}}}}
\newcommand{\OXt}{\O_{\text{\tiny {\it X}}}}
\newcommand{\rmp}{\r^{(m)}_{_0}}
\newcommand{\Omp}{\O^{(m)}_{_0}}
\newcommand{\rmf}{\r^{(m)}_{\text{\footnotesize eff}}}

\newcommand{\Omf}{\O^{(m)}_{\text{\footnotesize eff}}}
\newcommand{\Omft}{\O^{(m)}_{\text{\ssmall eff}}}

\newcommand{\sw}{\mathsf w}
\newcommand{\wJ}{\sw_{\!_J}}
\newcommand{\wx}{\sw_{\!_X}}

\newcommand{\Hp}{H_{_0}}
\newcommand{\fp}{\f_{_0}}
\newcommand{\tp}{t_{_0}}

\newcommand{\bmax}{\b_{\text{max}}}

\newcommand*\rfrac[2]{{}^{\displaystyle #1}\!/_{\! \displaystyle #2}}
\newcommand*\rfra[2]{{}^{\scriptstyle{#1}}\!\!\diagup_{\!\!\scriptstyle{#2}}}
\newcommand*\rfraa[2]{{}^{\displaystyle #1}\!\!\!\diagup_{\!\!\displaystyle #2}}
\newcommand*\rfrr[2]{{}^{#1}\!/_{#2}}

\newcommand{\bdm}{\begin{displaymath}}
\newcommand{\edm}{\end{displaymath}}

\long\def\symbolfootnote[#1]#2{\begingroup%
\def\thefootnote{\fnsymbol{footnote}}\footnote[#1]{#2}\endgroup}
\numberwithin{equation}{section}
\providecommand{\appendixname}{appendix}

\begin{document}

\markboth{Arshdeep Singh Bhatia and Sourav Sur}
{Dynamical system analysis of dark energy models ...}

%
%

\title{DYNAMICAL SYSTEM ANALYSIS OF DARK ENERGY MODELS IN SCALAR COUPLED METRIC-TORSION THEORIES}

\author{\large Arshdeep Singh Bhatia$^*$ and Sourav Sur$^\dagger$}

\address{{\normalsize \it Department of Physics \& Astrophysics\\ 
University of Delhi\\
New Delhi - 110 007, India}\\ \vskip0.05in
{\it $^*$arshdeepsb@gmail.com; asbhatia@physics.du.ac.in\\
$^\dagger$sourav.sur@gmail.com; sourav@physics.du.ac.in}}
\maketitle

\begin{abstract}
We study the phase space dynamics of cosmological models in 
the theoretical formulations of non-minimal metric-torsion 
couplings with a scalar field, and investigate in particular 
the critical points which yield stable solutions exhibiting 
cosmic acceleration driven by the {\em dark energy}. The 
latter is so defined that it effectively has no direct 
interaction with the cosmological fluid, although in an 
equivalent scalar-tensor cosmological setup the scalar field 
interacts with the fluid (which we consider to be the 
pressureless dust). Determining the conditions for the 
existence of the stable critical points we check their 
physical viability in both Einstein and Jordan frames. We 
also verify that in either of these frames, the evolution 
of the universe at the corresponding stable points matches 
with that given by the respective exact solutions we have 
found in an earlier work (arXiv:1611.00654 [gr-qc]). We not 
only examine the regions of physical relevance in the phase 
space when the coupling parameter is varied, but also demonstrate 
the evolution profiles of the cosmological parameters of interest 
along fiducial trajectories in the effectively non-interacting 
scenarios, in both Einstein and Jordan frames. 
\end{abstract}


\section{Introduction}  \label{sec:intro}

Dynamical stability is a major requirement for cosmological solutions representing {\em 
dark energy} (DE) that supposedly drives the late-time cosmic acceleration \cite{acc}. 
While the question as to how the DE evolves has been contemplated by a plethora of 
theoretical surmises and conjectures \cite{cope,derev,wols}, observations have mostly 
been in favour of a {\em non-dynamical} DE, reminiscent of a {\em cosmological constant} 
$\L$, at low to moderately high redshifts \cite{wmap,jla,planck}. However, some scope is 
there to look for (albeit mild) deviations from the concordant $\L$CDM model, comprising 
of $\L$ and cold dark matter (CDM) as the dominant constituents of the universe 
\cite{cc,lcdm,effcc}. In fact, the dynamical aspects of the DE are always worth examining, 
for a sufficiently longer span of evolution, tracing back from deep in the past, till 
extrapolating to high blueshifts in the future \cite{derev,wols,tsuj}. The theoretical 
motivation for this is obvious, in view of the well-known {\em fine tuning} and 
{\em coincidence} problems affecting the $\L$CDM cosmology \cite{cc,lcdm}.

Extensive searches for the dynamical DE, within the standard Friedmann-Robertson-Walker 
(FRW) framework, have mostly accounted for the scalar field candidates, such as quintessence, 
k-essence, tachyon, dilaton, chameleon, etc. \cite{quin,kess,tach,dil,cham}, which have had 
many intriguing features \cite{cope,derev}. However, in recent years the focus has shifted to 
a purely geometric characterization of the DE in the so-called {\em modified gravity} theories
\cite{modgrav} of e.g. the $f(\cR)$ type \cite{fR}, where $\cR$ is the Riemannian curvature 
scalar. Such theories can also be mapped to scalar-tensor theories \cite{BD,frni,fujii,scaltens}, 
and hence give rise to interacting (or unified) dark energy--matter scenarios \cite{int-de} 
under conformal transformations. One's perception though, of a `geometrical' DE, is not limited 
to the formulations in the Riemannian space-time only. We may equally well look into the 
cosmologies emerging from the rather conventional extensions of General Relativity (GR), such
as that formulated in the four-dimensional {\it Riemann-Cartan} ($U_4$) space-time  with {\em 
torsion} --- an antisymmetric tensor field that generalizes the Levi-Civita connections in GR
\cite{hehl,akr,traut,sab,shap,west,ssasb}. Torsion is often considered as a geometric entity 
that provides a classical background for quantized {\it spinning} matter, and is therefore an 
inherent part of a fundamental (quantum gravitational) theory, such as string theory
\cite{shap,quantor,ham}. A completely antisymmetric torsion can have its source in the closed 
string massless Kalb-Ramond mode \cite{pmssg,saa}, with interesting implications in cosmology
and astrophysics \cite{ssgss,skpm,rubhehl,dasetal,dmpmssg,acpm,bmssgsen,ssgss1,bmssgss,scssg}. 
Among other torsion scenarios of interest in the cosmological context, most notable are those 
based on the {\em teleparallel} $f(T)$ theories \cite{fT}, extended gravity theories
\cite{extgrav}, Poincar\'e gauge theory of gravity \cite{yonest,mink,hehlblag}, etc.

We in this paper turn our attention to the formalism of a metric-scalar-torsion (MST) theory 
developed in an earlier work (henceforth `paper I') \cite{ssasb2}. Such a theory deals with 
the $U_4$ Lagrangian non-minimally coupled to a scalar field $\f$ (of presumably primordial 
origin), in a way that no uniqueness problem arises \cite{shap,ssasb2,neto}. Now, in the 
standard cosmological framework, the torsion degrees of freedom get restricted by the FRW 
metric structure \cite{ssasb}. Also since $\f$ acts as the source of the trace mode of 
torsion (via the corresponding equation of motion), we effectively have a scalar-tensor 
equivalent MST setup. The pseudo-trace mode of torsion can give rise to a mass term for 
$\f$, via suitable augmentation of the effective action with say, some higher order torsion 
terms \cite{ssasb2}. Considering further a pressureless {\em dust}-like cosmological matter, 
viable DE solutions in Einstein and Jordan frames have been worked out analytically in paper 
I \cite{ssasb2}, keeping the cosmological parameters within the corresponding error estimates
for the $\L$CDM model from recent observations. However, there remains the important question:

\vskip 0.05in
\no 
``{\em would these (and possibly a few other) solutions, persist over time (i.e. stable), 
once subjected to fluctuations in the solution space (or the phase space)}?"

\vskip 0.05in
\no  
Answering this requires an in-depth analysis of the MST-cosmological dynamics in both Einstein 
and Jordan frames. Our objective in this paper is to carry out such an analysis, by constructing 
(from cosmological equations in either cases) the autonomous system of equations in terms of 
suitable phase space variables.

For simplicity, we take into account only the two dominant components of the universe, viz. the 
dust and the scalar field $\f$, whence the phase space is a two-dimensional (i.e. a {\em phase 
plane}). However, instead of working with $\f$ and its mass $m$, throughout the analysis we resort 
to a {\em torsion scalar} $\t \le(\sim 3 \ln \f\ri)$ and a {\em torsion constant} $\L \le(\sim 
m^2\ri)$, so as to have a clear understanding of torsion's effect on the dynamics. Whereas $\t$ 
is equal to the time-integral of the norm of the torsion trace vector (as defined in the original 
Jordan frame), $\L$ is given by the norm of the pseudo-trace vector of torsion (modulo some 
numerical factor).

We follow the standard methodology based on linear perturbation theory \cite{cope,tsuj} to determine 
the {\em critical points} (CPs) in the phase plane and their characteristic type and nature. This 
is essential, since each CP represents an equilibrium state of the system (the universe) in the 
asymptotic limit ($N = \ln a \rightarrow \infty$, where $a$ is the cosmological scale factor). Now,
a given cosmological solution is considered {\em stable} if it transpires to the dynamical evolution 
of the universe at a stable CP. However, as is common in a plethora of contexts in the literature
\cite{cope,tsuj,sprt,huang,chengong}, there are instances of more than one CPs existing in a certain 
{\it parametric domain}, i.e. the range of values of a system parameter (e.g. a coupling parameter). 
The same is the situation we find here, in both Einstein and Jordan frames, for certain domains of 
our MST-coupling parameter $\b$. This compels us to analyse the dynamical evolution at each individual 
CP and figure out the appropriate one(s) in the respective (Einstein or Jordan) frame. Additionally, 
we have to determine the parametric domain(s) in which a stable CP supports solution(s) that exhibit
cosmic acceleration in the asymptotic limit. We do so in both Einstein and Jordan frames, and hence 
show that the corresponding exact solutions found analytically in paper are indeed stable. Numerically
solving the autonomous equations, subject to appropriate initial conditions, we work out a host of
relevant trajectories in the phase plane, in order to examine the overall dynamics of the system and 
its constituents leading up to the CPs. Also, for certain fiducial settings, we demonstrate the 
evolution of cosmological parameters of interest, such as the effective DE density and equation of 
state (EoS) parameters, $\OX$ and $\wx$, along the corresponding trajectories. 

We carry out the dynamical analysis first in the Einstein frame, in which the cosmological 
equations are rather simple and have resemblance with those for quintessence. There are
however two major differences. Firstly, the torsion scalar $\t$ interacts with the (apriori 
dust-like) fluid, thus affecting the dynamics of both. As such, it is not possible to make 
a direct comparison of the MST-cosmological parameters with those estimated (from observations) 
for known models, such as $\L$CDM. It is rather convenient to resort to the scenario in which 
the critical density of the universe, $\r$, is decomposed into two effective non-interacting 
components, viz. the dust-like matter and a left-over, supposedly the DE \cite{ssasb2}. In 
such a scenario, the physical relevance of the existent CPs in the phase plane is implicated 
by the eventual extinction of the matter sector, irrespective of the initial conditions. 
Secondly, it is not desirable to have the chosen phase space variables depending explicitly 
on the system parameter $\b$. Otherwise their calibration would keep on changing with $\b$, 
different domains of which are assigned for the existence of the CPs and(or) their physical 
relevance. Therefore, if instead of $\t$ we choose to work with a redefined (quintessence-like) 
field $\vph$ which absorbs $\b$ in it, the {\em physically admissible regions} $\fM$ for the 
trajectories in the phase plane would get altered in shape and size \cite{ssasbPP}. 

Repeating the analysis in the Jordan frame is straightforward, but cumbersome because of an
explicit $\t$-dependence of the gravitational coupling factor $\keff^2 \equiv 8 \pi \Geff$,
where $\Geff$ is the generalization of the Newton's constant. There are some interesting 
consequences of this though, culminating from two legitimate standpoints. In principle, we 
may resort to {\it one} of the following: (a) a conventional scenario in which the critical 
density $\rJ$ varies with $\keff$ and is not conserved, although the matter density is 
conserved \cite{tsuj}, and (b) an effective scenario in which the critical density $\r$, 
defined as in a minimally coupled theory, is the sum of the densities of the dust and a 
left-over (supposedly the DE), which are individually conserved \cite{ssasb2}. Now, since 
it is the same Jordan frame MST setup looked from different perspectives, the general 
outcomes of the dynamical analysis remain the same in both the scenarios, viz. the same 
number of CPs of the same type and nature. One difference is there though --- in the 
conventional scenario, the physically admissible regions $\fM$ are confined within two 
similar curves (conic sections) in the phase plane, whereas only the outer curves are there 
in the effective scenario, for the same values of the effective Brans-Dicke parameter $\fw$ 
($\propto \b^{-1}$)\footnote{We choose to take $\fw$ as the Jordan frame system parameter, 
different domains of which ascribe to the existent and(or) the physical relevant CPs. The 
limiting values set on $\fw$ from extensive studies \cite{BDbounds1,BDbounds2}, provide an 
independent credibility check of the cosmological solutions \cite{ssasb2}.}. The cosmological 
dynamics in the effective scenario, although by-and-large similar to that in the Einstein 
frame, has an intriguing feature, viz. the existence of a stable CP that supports solutions 
which not only exhibit cosmic acceleration in the asymptotic limit, but also a {\em 
super-accelerating} or {\em phantom} regime in course of their evolution. One such stable 
solution is actually that found in paper I, for which the phantom barrier crossing takes
place at an epoch in the near past, whereafter the phantom regime continues eternally 
\cite{ssasb2}.

This paper is organized as follows: starting with a general description of the MST formalism 
in \S \ref{sec:mst}, we write down the scalar-tensor equivalent actions in both Jordan and 
Einstein frames, in terms of the torsion scalar $\t$. Considering first the Einstein frame 
MST-cosmological setup in \S \ref{sec:E-PP}, we proceed as follows: (i) work out (in \S 
\ref{sec:E-eqs}) the equations of motion for the effectively non-interacting dust and DE 
sectors, (ii) construct from them (in \S \ref{sec:E-aut}) the autonomous system of equations 
and examine the domains of existence and(or) physical relevance of the CPs and also their 
type and nature, (iii) study (in \S \ref{sec:E-dyn}) the dynamical evolution of the universe 
at each of the CPs and examine the stability of viable DE solutions such as the one found in 
paper I, and finally (iv) obtain the phase plane trajectories (in \S \ref{sec:numaut}) by 
numerically solving the autonomous equations for appropriate sets of initial conditions, and 
hence illustrate the evolution of the DE density and EoS parameters along a fiducial trajectory 
that leads up to a stable CP. Almost the same chronology is maintained while repeating the
dynamical analysis in \S \ref{sec:J-PP} for the MST-cosmological setup in the Jordan frame. 
Some characteristic differences are there of course, compared to the Einstein frame analysis. 
Accordingly, a detailed account of, for e.g. the shape and size of the physically admissible 
region(s) $\fM$ with the variation in the system parameter $\fw$, is given. Also the phantom
barrier crossing in the effective Jordan frame scenario is illustrated clearly, by working 
out the evolution along trajectories for fiducial settings corresponding to two different 
values of $\fw$. We conclude in \S \ref{sec:concl} with a summary and a discussion on some 
implications and possible extensions. 

We use the same notations and conventions as in paper I, viz. the metric signature throughout 
is $\le(-,+,+,+\ri)$, units are chosen so that the speed of light $c = 1$, and the determinant 
of the metric tensor $g_{\m\n}$ is denoted by $g$.

\section{The general MST formalism in the cosmological setup}  \la{sec:mst}

Let us discuss the basic tenets of the metric-scalar-torsion (MST) formalism, viz. that of 
scalar field couplings to the four-dimensional Riemann-Cartan ($U_4$) Lagrangian \cite{ssasb2}. 
The $U_4$ space-time is characterized by an asymmetric affine connection: $\Ct^{\l}_{~~\m\n} (\neq
\Ct^{\l}_{~~\n\m})$, which incorporates the third-rank {\em torsion} tensor defined as: 
$T^{\l}_{~~\m\n} := \Ct^{\l}_{~\m\n} - \Ct^{\l}_{~\n\m} \le(= -\, T^{\l}_{~~\n\m}\ri)$. 
Essentially, the Riemannian ($R_4$) covariant derivative $\nab_{\m}$ in GR (defined via the 
symmetric Levi-Cevita connections $\C^{\l}_{~\m\n}$) is replaced with that ($\nt_{\m}$) 
defined via $\Ct^{\l}_{~\m\n}$, preserving the metricity condition $\nt_{\m} g_{\a\b} = 0$. 

The torsion tensor can be decomposed into three irreducible modes, viz. the trace $\, \cT_\m 
:= T^\n_{~\m\n}$, the pseudo-trace $\, \cA^\s := \e^{\a\b\c\s} \,T_{\a\b\c}$ and the
(pseudo-)tracefree tensorial residue $\, \cQ_{\m\n\s}$, whence the $U_4$ analogue of the 
Ricci scalar curvature $\cR$ is 
\be \la{U4-curv}
\Rt \,:=\, \cR \,-\, 2 \, \nab_\m \, \cT^\m \,-\, \fr 2 3 \, \cT_\m \cT^\m 
\,+\, \fr 1 {24} \, \cA_\m \cA^\m \,+\, \fr 1 2 \cQ_{\m\n\s} \cQ^{\m\n\s} \,\, .
\ee
Accordingly, the free $U_4$ Lagrangian $\, L_{U_4} = \sq{- g} \, \Rt$ has a purely algebraic 
dependence on torsion\footnote{Note that the $\sq{- g} \, \nab_\m \, \cT^\m$ term in $L_{U_4}$ 
is merely a total divergence (or, a boundary term).}. Now, while coupling a scalar field $\f$ 
to $L_{U_4}$, one encounters the well-known problem of {\it non-uniqueness} of the resulting
action under the minimal coupling scheme ($\pa_\m \rightarrow \nt_\m$) \cite{shap,buchshap}. A 
simple (and convenient) way to avoid this is to assume a non-minimal term $\f^2 \, L_{U_4}$, 
so that upto a total divergence the action is \cite{ssasb2}
\bea \la{MST-ac}
\cS = \int d^4 x \sq{-g} \le[\fr{\b \f^2} 2 \le(\cR + 4 \cT^\m \fr{\pa_\m \f}{\f} 
- \fr 2 3 \cT_\m \cT^\m + \fr 1 {24} \cA_\m \cA^\m + \fr 1 2 \cQ_{\m\n\s} 
\cQ^{\m\n\s}\ri) \ri. ~~~ \nn\\
\le. - \fr 1 2 \, g^{\m\n} \, \pa_\m \f \, \pa_\n \f - V (\f) + \cL^{(m)} \ri] , ~~~ 
\eea
where $\b$ is a dimensionless coupling constant, $V (\f)$ is the scalar field potential, and 
$\cL^{(m)}$ is the Lagrangian density for other matter fields in the theory. Eq. (\ref{MST-ac}), 
dubbed as the `MST action' \cite{ssasb2}, leads to the equation of motion $\, \cT_\m = 
\le(\rfraa{3 \,}{\! \f}\ri) \pa_\m \f$, which implies that the scalar field $\f$ acts a source 
of the trace mode $\cT_\m$ of torsion. Moreover, in order to preserve the FRW metric structure 
in a standard cosmological setup, one requires the tensor mode $\cQ_{\m\n\s}$ of torsion to 
vanish altogether, and the vector modes $\cT_\m$ and $\cA^\m$ to have only their temporal 
components existent \cite{ssasb,tsim}. Also since the torsion field is generally taken to be 
massless \cite{hehl,ham}, one expects the scalar field source $\f$, of its trace mode $\cT_\m$ 
to be massless as well. However, the pseudo-trace mode of torsion, $\cA^\m$, may effectively 
lead to a scalar field potential $V (\f) = \rfraa{m^2 \f^2} 2$, where $m$ is a mass parameter 
for $\f$. Such a possibility arises from a suitable augmentation of the MST action (\ref{MST-ac}) 
with say, some higher order torsion term(s) such as $\, \f^2 \! \le(\cA_\m \cA^\m\ri)^2$, whence
one gets $\, \cA_\m \cA^\m =\, - \, \rfraa{48 \, m^2\!} \b$ via the corresponding equation of 
motion\footnote{Note that, a mass term for $\f$ may also result from a {\em norm-fixing} 
constraint on $\cA^\m$ \cite{ssasb2}, similar to that in the vector-tensor gravity theories of 
Einstein-{\ae}ther type \cite{jacob,carroll,gao}, or in the mimetic gravity theories 
\cite{dde,mimgrav,vag}. However, such an analogy has no specific physical motivation.} 
\cite{ssasb2}. The effective MST action then assumes the form of a scalar-tensor action in the 
{\em Jordan} frame \cite{ssasb2}:
\be \la{MST-ac2}
\cS = \int d^4 x \, \sq{-g} \le[\le(\!\fr{\f}{\fp}\!\ri)^{\!\! 2} \! \fr{\cR}{2 \k^2} \,-\, 
\fr{\le(1 - 6\b\ri)} 2 \, g^{\m\n} \, \pa_\m \f \, \pa_\n \f \,-\,\fr 1 2 \, m^2 \f^2 \,+\, 
\cL^{(m)}\ri] \,,
\ee
where $\k = \sq{8 \pi G}$ and $\fp = \le(\k \sq{\b}\ri)^{-1}$ is the value\footnote{The parameter 
$\b$ is of course taken to be positive definite, as otherwise the underlying quantum gravitational 
theory would be unbounded from below.} of $\f$ at the present epoch $\tp$, such that the running gravitational coupling parameter $\, \Geff \sim \f^{-2}$ has its present-day value $\, \Geff (\tp) 
= G$, the Newton's constant. 

Let us now define a dimensionless scalar field $\t$ as:
\be \la{tfield}
e^\t := \le(\rfraa{\f}{\fp}\!\ri)^{\!\! 3} \,,
\ee
so that at $t = \tp$, $\t = 0$, and the effective gravitational coupling factor is given by
\be \la{kappa-eff}
\keff (\t) \,\equiv \sq{8 \pi \, \Geff (\t)} \,=\, \k \, e^{\rfra{- \t} 3} \,\,.
\ee
The Jordan frame action (\ref{MST-ac2}) is then expressed as
\be \la{J-ac}
\cS \,=\, \int d^4 x \, \sq{-g} \le[e^{\rfra{2 \t \!} 3} \le\{\fr{\cR}{2 \k^2} \,-\, \fr \vep 2 \,
g^{\m\n} \, \pa_\m \t \, \pa_\n \t \,-\, \L\ri\} +\, \cL^{(m)}\ri] \,,
\ee
where $\vep$ and $\L$ are two dimensionful constants, given by
\be \la{Ldef}
\vep \,=\, \fr{\le(1 - 6 \b\ri) \fp^2} 9 \,=\, \fr{1 - 6 \b}{9 \k^2 \b}  \qquad \mbox{and}
\qquad \L \,=\, \fr 1 2 \, m^2 \fp^2 \,=\, \fr{m^2}{2 \k^2 \b} \,\,.
\ee
One may note that the field $\t$ and the constant $\L$ facilitates a clear understanding of 
the roles of the individual torsion modes in the MST-cosmological dynamics. If one makes 
interpretations in terms of torsion parameters chosen as the norms of the trace and 
pseudo-trace vector modes of torsion \cite{ssasb2}, then it is easy to see that
\be \la{TAnorm}
\le|\cT\ri| := \sq{- g^{\m\n} \cT_\m \cT_\n} = \sq{- g^{\m\n} \pa_\m \t \pa_\n \t} 
\quad \mbox{and} \quad 
\le|\cA\ri| := \sq{- g^{\m\n} \cA_\m \cA_\n} = 4 \k \sq{3 \L} \,.
\ee
So, we always have $\L \propto \le|\cA\ri|^2$, and simply $\, \t = \int dt \le|\cT\ri| \,$ in 
a cosmological space-time described by the spatially flat FRW metric, viz. $\, g_{\m\n} =$
diag$[-1, \, a(t), \, a(t), \, a(t)]$, where $t$ is the comoving time and $a(t)$ is the scale 
factor. Henceforth, we shall appropriately refer to $\t$ as the `torsion scalar' and $\L$ as 
the `torsion constant'.

Now, as is usual in a scalar-tensor equivalent theory, the equations of motion are much simpler 
in the {\em Einstein} frame than in the Jordan frame \cite{frni,fujii}. The Einstein frame 
MST-action can be obtained from Eq. (\ref{J-ac}), under the conformal transformation $\, 
g_{\m\n} \lra \hg_{\m\n} = e^{\rfra{2 \t \!} 3} \, g_{\m\n}$ \cite{ssasb2}:
\be \la{E-ac}
\hS \,=\, \int d^4 x \, \sq{-\hg} \le[\fr{\hR}{2 \k^2} \,-\, \fr {\zeta^2} 2 \, \hg^{\m\n} \, 
\pa_\m \t \, \pa_\n \t \,-\, \L \, e^{\rfra{- 2 \t \!} 3} \,+\, \hLm \!\le(\hg,\t\ri) \ri] \,,
\ee
where $\zeta = (3 \k \sq{\b})^{-1}$ is a dimensionful constant, $\hg \equiv \text{det}
\le(\hg_{\m\n}\ri)$ is the Einstein frame metric determinant, $\hR$ is the corresponding 
curvature scalar, and $\, \hLm \!\le(\hg,\t\ri) = e^{\rfra{- 4 \t \!} 3} \, \cL^{(m)} 
\!\le(g (\hg, \t)\ri)$ is the corresponding matter Lagrangian density. Despite their mathematical
equivalence, the Einstein and Jordan frames in general have different outcomes of physical 
measurements. The reason is obviously the gravitational coupling factor, which varies in one 
frame and not in the other. In fact, there is a longstanding debate as to which of these frames 
is actually of physical relevance \cite{frni,fujii}. For completeness therefore, we shall 
subsequently carry out the dynamical analysis for our MST-cosmological formalism in both 
the Einstein and Jordan frames, taking one or the other to be physically relevant. We shall set 
up first the corresponding (Einstein or Jordan frame) cosmological equations, for the two system
constituents, viz. the torsion scalar $\t$ and the cosmological matter in the form of a 
pressureless (non-relativistic) {\em dust}. Defining suitable variables for the corresponding 
two-dimensional phase space (or the {\em phase plane}), we shall then construct the autonomous 
system of equations and look for stable solutions representing an effective DE evolution. Since 
the dust {\it feels} the effect of torsion\footnote{That is, the dust-like fluid and the torsion 
scalar $\t$ have a mutual interaction, which results from either the conformal transformation or 
the varying gravitational coupling factor $\keff (\t)$.}, we shall resort to an effectively 
non-interacting picture in the respective (Einstein or Jordan) frame. 

\section{Phase plane analysis in the Einstein frame} \la{sec:E-PP}

Let us consider, in this section, the Einstein frame to be suitable for physical observations, 
and drop for brevity the hats over all quantities defined in this frame. We shall however 
continue with the expressions (\ref{TAnorm}) for the norms $\le|\cT\ri|$ and $\le|\cA\ri|$, as 
defined in the Jordan frame, in order to keep track of the individual terms of our original 
MST-action (\ref{MST-ac}). We have therefore the relationships:
\be \la{E-tau}
e^{\rfra{\t} 3} \,=\, \fr 1 3 \int dt \, \le|\cT\ri| \qquad \mbox{and} \qquad
\L \,=\, \fr{\le|\cA\ri|^2}{48 \, \k^2} \,\,,
\ee
in the Einstein frame, with $t$ being the corresponding comoving time coordinate.

\subsection{Cosmological equations and the effective scenario} \la{sec:E-eqs}

The Friedmann and Raychaudhuri equations, obtained from the action (\ref{E-ac}), are
\be \la{E-eq}
H^2  \,=\, \fr{\k^2} 3 \le[\rmt \,+\, \rtt\ri]  \qquad \mbox{and} \qquad
\dot{H} \,=\, - \, \fr{\k^2} 2 \le[\rmt \,+\, \rtt \,+\, \ptt\ri] \,,
\ee
where $\, H := \rfraa{\dot{a}} a$ is the Hubble parameter corresponding to the Einstein frame
scale factor $a (t)$ (the overhead dot $\{\cdot\} \equiv \rfraa d {dt}$), $\, \rmt$ is the 
energy density of the fluid matter, whereas $\rtt$ and $\ptt$ are respectively the field energy 
density and pressure:
\be \la{E-f-denpres}
\rtt \,=\, \fr{\zeta^2} 2 \, \dot{\t}^2 \,+\, \L \, e^{\rfra{- 2 \t \!} 3} 
\qquad \mbox{and} \qquad
\ptt \,=\, \fr{\zeta^2} 2 \, \dot{\t}^2 \,-\, \L \, e^{\rfra{- 2 \t \!} 3} \,\,.
\ee
The corresponding energy-momentum conservation relation, given by
\be \la{E-consv}
\dot{\r}^{(m)} + 3 H \rmt = - \rmt \fr{\dot{\t}} 3 
\qquad \mbox{and} \qquad
\dot{\r}^{(\t)} + 3 H \le(\rtt + \ptt\ri) = \rmt \fr{\dot{\t}} 3 \,\,, 
\ee
imply that the cosmological fluid does not retain its `dust' interpretation in the Einstein 
frame, as its energy density depends explicitly on the torsion scalar $\t$:
\be \la{E-matdens}
\rmt (t) \,=\, \fr \rmp {a^3 (t)} \, e^{\rfra{- \t (t) \!} 3} \,\,,
\ee
where $\, \rmp = \rmt \rvert_{t=\tp} = \rmt \rvert_{a=1}$ is the present-day value of $\rmt$.

Now, it is easy to see that under a field redefinition $\zeta \t (t) \equiv \vph (t)$, the 
above Eqs. (\ref{E-eq})--(\ref{E-matdens}) correspond to those in an interacting system 
of an apriori pressureless cosmological fluid and a quintessence scalar field $\vph$ with 
an exponential potential \cite{quin}. Such a correspondence is however misleading, since 
the dynamical analysis of the system in terms of the torsion scalar $\t$ has a radical
difference with that in terms of the redefined field $\vph$ (see the discussion in the next 
subsection). In other words, working with $\t$ not only pinpoints the dynamical effects of
torsion on the evolution of the universe, but also leads to results different from those 
of the dynamical analysis for the standard scalar-tensor cosmologies in the Einstein frame
(which are essentially the systems of interacting quintessence and cosmological matter) 
\cite{tsuj}. Moreover, Eq. (\ref{E-matdens}) suggests that the expression for the {\em 
critical} (or {\em total}) density of the universe, viz. $\r = \rfraa{3 H^2 \!\!}{\! \k^2} 
= \rmt + \rtt$, is barely of any use when it comes to making a comparison with the parametric 
estimations of well-known models, such as $\L$CDM, from physical observations. It is rather 
convenient to express \cite{ssasb2}
\be \la{E-crit}
\r \,:=\, \fr{3 H^2}{\k^2} \,=\, \rmf \,+\, \rx \,\,,
\ee
where $\, \rmf$ is an effective (dust-like) matter density and $\, \rx$ is a surplus density 
(which we consider to be due to the DE). These are given respectively as 
\be \la{E-effdens}
\rmf \,=\, \fr{\rmp}{a^3} \,=\, \rmt \, e^{\rfra{\t \!} 3} \,\, 
\qquad \mbox{and} \qquad
\rx \,= \rtt \,+ \le(e^{\rfra{- \t \!} 3} \,-\, 1\ri) \rmf \,\,.
\ee
The Friedmann equation can then be recast as
\be \la{E-Fried}
\Om \,+\, \Otau \,=\, \Omf \,+\, \OX \,=\, 1 \,\,,
\ee
where $\, \Om = \rfraa{\rmt\!}{\r}$ and $\, \Omf = \rfraa{\rmf\!}{\r}$ are the actual and the effective 
matter density parameters respectively, $\, \Otau = \rfraa{\rtt\!}{\r}$ is the density parameter for 
the field $\t$, whereas $\, \OX = \rfraa{\rx}{\r}$ is that for the DE. Identifying the DE pressure as 
$\, \px = \ptt$, we also have the DE conservation relation (obtained using Eqs. (\ref{E-consv})):
\be \la{E-DEconsv}
\dot{\r}_{_X} \,+\, 3 \, H \le(\rx \,+\, \px\ri) =\, 0 \,\,.
\ee
Eqs. (\ref{E-effdens})--(\ref{E-DEconsv}) govern the dynamical evolution of the system. One may 
in principle look for their outright solutions, for e.g. by guessing suitable solution ansatze, 
and then examine the physical viability of those solutions \cite{ssasb2}. A rather general 
alternative is to construct an autonomous system of first order coupled differential equations, 
out of Eqs. (\ref{E-effdens})--(\ref{E-DEconsv}), and look for such a system the plausible real 
roots and their stability against small fluctuations in the solution space (or the phase space).

\subsection{Autonomous equations and the critical points} \la{sec:E-aut}

Defining the phase space variables as
\be \la{E-aut}
X \,:=\, \fr{\dot{\t}} {3 \sq{6} H}  \qquad \mbox{and} \qquad
Y \,:=\, \fr{\k \sq{\L} \, e^{\rfra{- \t \!} 3}}{\sq{3} H} \,\,,
\ee 
we obtain from the above cosmological equations, the autonomous equations:
\bea
&& \fr{dX}{dN} \,=\, \fr 3 2 \le(X - \b \sq{\fr 2 3}\ri) \le(\fr{X^2}{\b} - Y^2 - 1\ri) \,\,, 
\la{E-auteq1} \\
&& \fr{dY}{dN} \,=\, \fr{3 Y} 2 \le(\fr{X^2}{\b} - Y^2 - 2 \sq{\fr 2 3} \, X + 1\ri) \,\,, 
\la{E-auteq2}
\eea
where $\, N (t) \equiv \ln a (t)$ is the number of e-foldings, and one also has the constraint 
\be \la{E-autcons}
\fr{X^2}{\b} \,+\, Y^2 \,-\, 1 \,+\, \Om \,=\, 0 \,\,.
\ee
Inverting the definition of the variable $X$ in Eq. (\ref{E-aut}), we express
\be \la{tauF}
\t (N) \,=\, 3 \sq{6} \, F (N)  \,\,, \qquad \mbox{where} \qquad
F (N) \,\equiv\, \int_0^N X(\cN) \, d\cN \,\,.
\ee
Eqs. (\ref{E-crit}), (\ref{E-effdens}) and (\ref{E-autcons}) imply that the effective matter 
density parameter is
\be \la{E-effdens1}
\Omf \,:=\, \fr{\rmf}{\r} \,= \le(1 - \fr{X^2}{\b} - Y^2\ri) e^{\sq{6} \, F} \,\,.
\ee
Moreover, the total pressure being $\, p = \px = \ptt$, we have from Eqs. (\ref{E-crit}), 
(\ref{E-effdens}) and (\ref{E-aut}) the total equation of state (EoS) parameter of the 
system  given by
\be \la{E-eos}
\sw \,:=\, \fr p {\r} \,=\, \fr{X^2}{\b} - Y^2 \,\,,
\ee
i.e. the EoS parameter for the DE is 
\be \la{E-deeos}
\wx \,:=\, \fr{\px}{\rx} \,=\, \fr{X^2 - \b \, Y^2}{\b \, \OX} \,\,,
\ee
since $\, \sw = \OX \wx$, where $\, \OX = 1 - \Omf$, by Eq. (\ref{E-Fried}).

It is worth noting here that the autonomous system of equations (\ref{E-auteq1})--(\ref{E-autcons}) 
is symmetric under the interchange $Y \rightarrow -Y$, which means that we can restrict our analysis 
to the region $Y \geq 0$ of the $XY$ phase plane without loss of generality. Moreover, as mentioned 
in the previous subsection, the MST-cosmological equations (\ref{E-eq}) and (\ref{E-f-denpres}) 
correspond to those for quintessence, under the redefinition $\, \vph (t) \equiv \zeta \t (t)$. Such
a correspondence may {\it not} in general be reflected in the dynamical analysis though, when the 
(dimensionless) phase space variables are defined using $\vph$. Actually, in comparison to the 
standard (and even the interacting) quintessence scenarios, the MST setup has the intriguing aspect 
of the coupling parameter $\b$ playing a potentially active role in determining the viable cosmologies. 
So it is imperative to allow for a discrete alteration of the value of $\b$ in the dynamical analysis. 
However, for simplicity we may keep the other parameter in the theory, viz. $\L$, to remain {\it fixed}. 
Now in such a situation, while working with $\vph$ instead of $\t$, we cannot use the above definitions 
(\ref{E-aut}) of the phase space variables $X$ and $Y$, with a mere substitution of $\t$ by 
$\rfraa{\vph}{\zeta}$ therein. The reason is that $\zeta$ being proportional to $\b^{-\rfrr 1 2}$, 
such a substitution would mean $X$ and $Y$ explicitly dependent on $\b$. Therefore their calibration 
would change when the value of $\b$ is changed, thus giving rise to an ambiguity in the analysis. So, 
in terms of $\vph$ one has to define altogether different phase space variables, under the demand that 
they need to be free from any explicit $\b$-dependence\footnote{Their implicit dependence on $\b$ is 
not a worry though, as that would not affect their calibration}. Hence, we would have a different set 
of autonomous equations which may lead to a different dynamics of the same system, if we resort to the 
redefined field $\vph$, instead of persisting with the torsion scalar $\t$. 

Now, the objective of analysing autonomous equations, say $\rfraa{dX}{dN} = \cF (X, Y)$ and 
$\rfraa{dY}{dN} = \cG (X, Y)$, is to determine the {\em critical points}, or the equilibrium 
solutions, and consequently examine the type and nature of such solutions, i.e. their stability
in the two-dimensional phase space formed by $X$ and $Y$. Here, $\cF$ and $\cG$ are given
functions of $X$ and $Y$ for a particular system, for e.g. the right hand sides of Eqs. 
(\ref{E-auteq1}) and (\ref{E-auteq2}). By definition, a critical point (CP) is assigned 
coordinates $\le(X_c, Y_c\ri)$ at which $\cF$ and $\cG$ vanish. Now, for small changes 
$\le(\d X, \d Y\ri)$ about $\le(X_c, Y_c\ri)$, we have the following eigenvalue equation 
in the linear perturbation theory \cite{cope}:
\begingroup                                    
\renewcommand*{\arraystretch}{1.35}
\be \la{cp-eigeneq}
\fr d {dN}
    \begin{pmatrix}
      \d X \\
      \d Y
    \end{pmatrix}
 =\, \cM 
    \begin{pmatrix}
      \d X \\
      \d Y
    \end{pmatrix} \,, \qquad \mbox{where} \qquad 
\cM \,\equiv
\le[\begin{array}{cc}
\rfraa{\pa \cF}{\pa X} & \rfraa{\pa \cF}{\pa Y} \\
\rfraa{\pa \cG}{\pa X} & \rfraa{\pa \cG}{\pa Y}
\end{array} \ri]_{\le(X_c, Y_c \ri)} \!\!\!\!.    
\ee 
\endgroup
\no 
The eigenvalues $\m_1$ and $\m_2$ of the matrix $\cM$, i.e. their type (real, imaginary or 
complex) and their sign, determine whether the solutions for $\d X$ and $\d Y$ have modes 
that are exponentially growing or decaying with $N$, and as such assert the type and the 
nature of the CPs. Specifically, there are the following cases: 
\ben[(i)]
\item Real $\m_1$ and $\m_2$: (a) If they are both negative, then $\le(\d X, \d Y \ri) 
\rightarrow 0$ as $N \rightarrow \infty$, which means that all phase space trajectories in 
the vicinity would terminate at the CP $\le(X_c, Y_c\ri)$, i.e. the latter acts as a stable 
{\em nodal sink}, or an {\em attractor}. (b) If they are both positive, then the perturbations 
$\le(\d X, \d Y \ri)$ build up over time (or $N$), taking the trajectories away from the CP 
$\le(X_c, Y_c\ri)$, which therefore acts as an unstable {\em nodal source}. (c) If they are 
of opposite sign, then the CP $\le(X_c, Y_c\ri)$ is a {\em saddle point}, i.e. it acts as an 
attractor along a particular direction (viz. the attractor axis) and as an unstable point 
along the direction normal to that. In general, the trajectories may tend towards the saddle 
point, but eventually get repelled away. Only for the trajectories that start from somewhere 
on the attractor axis, proceed along that, and manage to reach the saddle point, the latter 
acts as an attractor. Otherwise, the saddle point is unstable in nature.
\item Imaginary $\m_1$ and $\m_2$: The trajectories would in general describe an ellipse, 
having at its center the CP $\le(X_c, Y_c\ri)$, which is stable and is called a {\em center}.
\item Complex $\m_1$ and $\m_2$: Depending on whether their real parts are negative or positive, 
the trajectories would spiral towards or away from the CP $\le(X_c, Y_c\ri)$, rendering the latter 
to be a stable {\em spiral sink} or an unstable {\em spiral source}.
\een
In the case of vanishing $\m_1$ and $\m_2$, the type and nature of a CP are {\em indeterministic} 
if we persist with the linear perturbation theory, which actually breaks down \cite{cope}.

For our autonomous set of equations (\ref{E-auteq1})--(\ref{E-autcons}), the above methodology
implies the existence of {\it five} distinct CPs (discounting for the multiplicity of course).
The coordinates $\le(X_c, Y_c\ri)$ of these CPs and the domains of their existence and physical 
relevance (given by appropriate range of values of the parameter $\b$) are listed in Table 
\ref{E-tab-roots}. 
\begin{table}[phtb]
\tbl{Einstein frame critical points and the parametric domains of their 
existence and physical relevance.}{
\begin{tabular}{@{}cccc@{}} \toprule
\multirow{2}{*}{CP:} & Location $\le(X_c, Y_c\ri)$  
& \multicolumn{2}{c}{~~~~~~ Parametric domains of ~~~~} \\
& in the phase plane & existence & physical relevance \\
\colrule
$E_1$: & $\le(- \sq{\b}, \, 0\ri)$ & $\b \in \le(0, \infty\ri)$ & $\b \in \le(0, \infty\ri)$ \\ 
\colrule
$E_2$: & $\le(\sq{\b}, \, 0\ri)$ & $\b \in \le(0, \infty\ri)$ & $\b \in \le(0, \infty\ri)$ \\
\colrule
$E_3$: & $\le(\sq{\dfrac 2 3} \b, \, 0\ri)$ & $\b \in \le(0, \infty\ri)$ 
& $\b \in \le\{\rfraa 3 2\ri\}$ \\
\colrule
$E_4$: & $\le(\sq{\dfrac 2 3} \b, \, \pm \sq{1 - \dfrac{2\b} 3}\ri)$ 
& $\b \in \le(0, \rfraa 3 2\ri]$ & $\b \in \le(0, \rfraa 3 2\ri]$ \\
\colrule
$E_5$: & $\le(\sq{\dfrac 3 2}, \, \pm\, \sq{\dfrac 3 {2\b} - 1}\ri)$ 
& $\b \in \le(0, \rfraa 3 2\ri]$ & $\b \in \le\{\rfraa 3 2\ri\}$ \\ 
\botrule
\end{tabular} 
\label{E-tab-roots} }
\end{table}
The eigenvalues $\m_1$ and $\m_2$ of the linear perturbation matrix $\cM$ at each CP, and 
also the type and nature of the CPs are shown in Table \ref{E-tab-evalues}. 
\begin{table}[phtb]
\tbl{Eigenvalues of the linear perturbation matrix $\cM$ at the critical points, and the type 
and nature of these points in the Einstein frame.}{
\centering
\renewcommand{\arraystretch}{1.25}
\begin{tabular}{@{}cccc@{}} \toprule
CP & Eigenvalues of $\cM$ at CP & For parametric range: & CP type (nature) \\
\colrule
$E_1$ & $\m_1 = \m_2 = 3 + \sq{6\b} $ & $\b \in \le(0, \infty\ri)$: & Nodal Source (Unstable) \\ 
\colrule
\multirow{3}{*}{$E_2$} & \multirow{3}{*}{$\m_1 = \m_2 = 3 - \sq{6\b}$} 
&  $\b \in \le(0, \rfraa 3 2\ri)$: & Nodal Source (Unstable) \\
&  & $\b \in \le\{\rfraa 3 2\ri\}$: & Indeterministic \\
&  & $\b \in \le(\rfraa 3 2, \infty\ri)$: & Nodal Sink (Stable) \\
\colrule
\multirow{2}{*}{$E_3$} & \multirow{2}{*}{$\m_1 = - \m_2 = \dfrac{3 - 2\b} 2$} 
& $\b \in \le(0, \rfraa 3 2\ri) \cup \le(\rfraa 3 2, \infty\ri)$: & Saddle point (Unstable) \\
& & $\b \in \le\{\rfraa 3 2\ri\}$: & Indeterministic \\
\colrule
\multirow{2}{*}{$E_4$} & \multirow{2}{*}{$\m_1 = \m_2 = - \le(3 - 2\b\ri)$} 
& $\b \in \le(0, \rfraa 3 2\ri)$: & Nodal Sink (Stable) \\
& & $\b \in \le\{\rfraa 3 2\ri\}$: & Indeterministic \\
\colrule
\multirow{2}{*}{$E_5$} 
& \multirow{2}{*}{$\m_1 = - \m_2 = - \sq{\dfrac 6 {\b}} \le(\dfrac{3 - 2\b} 2\ri)$} 
& $\b \in \le(0, \rfraa 3 2\ri)$: & Saddle point (Unstable) \\
& & $\b \in \le\{\rfraa 3 2\ri\}$: & Indeterministic \\
\botrule
\end{tabular}
\la{E-tab-evalues} }
\end{table}
The criterion for their physical relevance actually follows from Eq. (\ref{E-effdens1}) 
for the effective matter density parameter $\Omf$ and the constraint (\ref{E-autcons}). The 
presence of the exponential factor $e^{\sq{6} \, F}$ in Eq. (\ref{E-effdens1}), where $\, F (N) 
\equiv \int_0^N X(\cN) \, d\cN$, implies that $\Omf$ may keep on evolving with time (or $N$) even 
after the system reaches a CP. So the condition for a physically realistic matter density, viz.
$\Omf < 1$, may get violated at some epoch, leaving the corresponding cosmological model {\it 
unphysical} at that CP. The exception(s) though is(are) the scenario(s) in which $\Omf = 0$ in 
the asymptotic limit. From Eq. (\ref{E-autcons}) we find the corresponding (physically relevant) 
CP(s) to be on the circumference of an ellipse, whose center is at the origin of the phase plane:
\be \la{E-cp-exist}
\fr{X_c ^2}{\b} \,+\, Y_c ^2 \,=\, 1 \,\,.
\ee
The region enclosed by this ellipse (let us call it $\fM$) is therefore the {\em physically
admissible} region in the phase plane. The identification of this region should be emphasized 
as the key difference between the dynamical system analysis here for the Einstein frame 
MST-cosmology, and that for the standard (quintessence-type) scalar field models in which the 
energy densities due to the field and the cosmological fluid matter are individually conserved. 
See below the comparison in the next subsection.

\subsection{Dynamical evolution of the universe at each critical point} \la{sec:E-dyn}

For viable cosmologies, particularly from the perspective of the late-time cosmic acceleration (or 
of the DE), it is necessary to identify the supportive CP(s). As the criterion (\ref{E-cp-exist}) 
of the physical relevance of the CPs follows from the argument that $\Omf = 0$ (i.e. $\OX = 1$)
asymptotically, the cosmological solutions at each of the CPs are given entirely by the energy 
component with the density parameter $\OX$. Although we are choosing to call it the 'DE', it is
yet to be verified whether this component actually complies with an accelerated expansion of the 
universe at late times (preceded by a regime of a decelerated one). Let us therefore look into 
the characteristics of the CPs and the dynamical aspects of the universe at each CP:    
\bed
\item CP $E_1$: Exists for all values of the parameter $\b$ and is situated at the intersection of 
the ellipse (\ref{E-cp-exist}) and the negative $X$-axis. It acts an {\em unstable} nodal source, 
since the eigenvalues $\m_1$ and $\m_2$ of the linear perturbation matrix $\cM$ are real and positive. 
The equilibrium state of the solutions, implicated by this CP and given by the DE with density 
parameter $\OX = 1$, is of an extremely decelerated expansion of the universe, because the EoS 
parameter of the DE is $\wx = 1$, i.e. the DE behaves like a steep fluid.
\item CP $E_2$: Exists for all values of $\b$ and is at the intersection of the ellipse (\ref{E-cp-exist}) 
and the positive $X$-axis. It acts as an {\em unstable} nodal source when $\b < \rfra 3 2$ and as a 
{\em stable} nodal sink when $\b > \rfra 3 2$. The solutions at this CP, in either case, are given by a 
steep fluid-like DE, exhibiting an extremely decelerated expansion of the universe. For $\b = \rfra 3 2$ 
however, the type and nature of this CP are {\em indeterministic}, as both the eigenvalues $\m_1$ and 
$\m_2$ of $\cM$ vanish.
\item CP $E_3$: Exists for all values of $\b$ and lies on the $X$-axis, but inside the ellipse 
(\ref{E-cp-exist}) except for $\b = \rfra 3 2$. It acts as an {\em unstable} saddle point (of no 
physical relevance though) for $\b \neq \rfra 3 2$, whereas for $\b = \rfra 3 2$ it coincides with 
$E_2$, whence its type and nature are {\em indeterministic}.   . 
\item CP $E_4$: Exists whenever $\b \leq \rfra 3 2$ and is constrained to lie on the circumference 
of the ellipse (\ref{E-cp-exist}). Whereas for $\b = \rfra 3 2$ it coincides with $E_2$ and $E_3$ 
(i.e. its type and nature are {\em indeterministic}), for $\b < \rfra 3 2$ it acts as a {\em stable} 
nodal sink and leads to a DE dominated accelerated expansion of the universe if further $\b < \rfra 
1 2$ (as can be checked easily\footnote{From Eq. (\ref{E-eos}) we see that at this CP $E_4$, viz. 
$\le(\sq{\rfra 2 3} \b, \sq{1 - \rfra{2\b} 3}\ri)$, the total EoS parameter for the system is $\sw 
= -1 + \rfra{4\b} 3$. So, the acceleration condition $\sw < - \rfra 1 3$ implies $\b < \rfra 1 2$.}). 
This is in fact the only CP (among the five) that can support solution(s) in presence of a 
non-vanishing potential term for the torsion scalar $\t$, and hence the viable DE model(s).
\item CP $E_5$: Exists whenever $\b \leq \rfra 3 2$, but is neither on the $X$-axis nor on the 
ellipse (\ref{E-cp-exist}), unless for $\b = \rfra 3 2$, whence it coincides with $E_2, E_3$ 
and $E_4$ (i.e. its type and nature are {\em indeterministic}). For $\b < \rfra 3 2$ it acts 
as an {\em unstable} saddle point (which is physically irrelevant as well). 
\eed
Let us now make a comparison with the (saddle and stable) CPs found in the dynamical analysis 
for the standard scenario of (non-interacting) dust and quintessence scalar field $\vph$ 
with an exponential potential $\sim e^{\k \l \vph}$, where $\l$ is some numerical factor 
\cite{cope,tsuj}. Among the saddle point(s) that could exist, depending on the value of 
$\l$, there is one always at the origin of the phase plane. This point supports solutions 
which require $\vph$ to get obliterated asymptotically, leaving the dust as the sole 
constituent of the universe. Among the stable point(s) that could exist, there is one 
that supports solutions which, depending on $\l$, are given in the asymptotic limit 
either entirely by a non-dynamic DE component (i.e. a cosmological constant $\L$) or by 
a $\L$CDM configuration with the densities of $\L$ and the dust of the same order of 
magnitude. In contrast, the analysis of our Einstein frame MST-cosmological dynamics 
leads to, depending on the value of the parameter $\b$, the saddle points $E_3$ ($\forall 
\b \neq \rfra 3 2$) and $E_5$ ($\forall \b < \rfra 3 2$), and the stable point $E_2$ 
($\forall \b > \rfra 3 2$) or $E_4$ ($\forall \b < \rfra 3 2$). However, we additionally 
have the criterion (\ref{E-cp-exist}) that permits only those solutions for which the 
cosmological matter gets obliterated asymptotically. None of the saddle points ($E_3$ 
and $E_5$) satisfies this criterion though. Nevertheless, as demonstrated below, the CP 
$E_3$ that lies in the region $\fM$ enclosed by the ellipse (\ref{E-cp-exist}), has 
significance in {\it funnelling} physical trajectories towards the stable point $E_4$. 
The latter is of course the only CP which supports solutions exhibiting the cosmic 
acceleration in the asymptotic limit, and that too for $\b$ restricted up to a maximum 
value $\bmax = \rfra 1 2$. So the DE models in the Einstein frame MST setup are viable 
only for a {\it fixed} parametric range $0 < \b < \rfra 1 2$. This also suggests (from 
Table \ref{E-tab-roots}) that only four CPs are of practical importance, viz. two 
unstable points $E_1$ and $E_2$ at the intersections of the $X$-axis and the ellipse 
(\ref{E-cp-exist}), a saddle point $E_3$ on the $X$-axis and inside this ellipse, and 
a stable point $E_4$ on the circumference of the ellipse. 

As to the status of the exact solution we have found in paper I by explicitly solving 
the Einstein frame MST-cosmological equations \cite{ssasb2}, note that: 

\vskip 0.05in
\no 
(i) The small parametric bound $\bmax \sim 10^{-2}$, for the viability of an almost 
$\L$CDM-like DE model described by such a solution \cite{ssasb2}, is compatible with the 
rather loose bound $\bmax = \rfra 1 2$ we have obtained here from the dynamical analysis. 

\vskip 0.05in
\no 
(ii) The universe described by such a solution must transpire to the dynamics of the DE 
and the cosmological matter at the stable point $E_4$. This can be seen by working out 
that at $E_4$, i.e. at $\le(X_c = \sq{\rfra 2 3} \b, \, Y_c = \sq{1 - \rfra{2\b} 3}\ri)$, 
the torsion scalar is given by $e^\t = a^{6 \b}$, whence the expression for the Hubble
parameter:
\be \la{Hub-E4} 
H^2 = \fr{\k^2 \, \L}{3 - s} \, a^{- 2 s} \,\,, \quad \mbox{under the substitution:} 
\,\,\, s = 2 \b \,,
\ee
is precisely the same as the asymptotic (i.e. the $a \rightarrow \infty$ limiting) form 
of that we have had in paper I, while deriving the exact solution in the Einstein frame 
(see section 4.1 therein). The stability of such a solution is thus established.

Now, for a clear understanding of the qualitative aspects of the evolution of the universe 
leading up to the stable point $E_4$, let us refer back for convenience, to the original 
decomposition of the critical density into the densities of the (interacting) torsion 
scalar $\t$ and the cosmological fluid in the Einstein frame. Note the following:

\vskip 0.1in
\no 
{\bf I.} Any point on the $Y$-axis of the phase plane implicates a non-dynamic $\t$ (i.e. $\dot{\t}
= 0$, since $X = 0$), whose contribution to the total energy density of the universe, via the potential 
$\L e^{\rfra{- 2 \t} 3}$, has a {\it fixed} value. The all-important torsion mode is therefore the 
pseudo-trace $\cA^\m$, which is assumed to give rise to the potential, that acts a cosmological constant 
$\L$ (with the corresponding EoS parameter $\wx = -1$). Hence the overall configuration for a point on 
the $Y$-axis is that of $\L$CDM. Moreover, the constraint (\ref{E-autcons}) implies that the further 
such a point is from the origin, the greater is the contribution of torsion to the energy content of 
the universe (subject always to the condition $\rfraa{X^2\!\!}{\b} + Y^2 \leq 1$, however). The $\L$CDM 
trajectory hence shows the evolution along the $Y$-axis, upto the point $(0,1)$, i.e. the apex of the 
elliptic boundary (\ref{E-cp-exist}) of the physically admissible region $\fM$ in the phase plane. 

\vskip 0.1in
\no 
{\bf II.} A point anywhere except on the $Y$-axis of the phase plane represents a system 
configuration in which the torsion scalar $\t$ is dynamical. The extent of such dynamics is 
determined by the magnitude of $X$, or equivalently by the contribution of the trace mode 
$\cT_\m$ of torsion to the total energy content of the universe. However, the viability of 
a DE evolution depends on how dominant is the potential $\L e^{\rfra{- 2 \t} 3}$, and hence 
the torsion pseudo-trace $\cA^\m$, over the dynamical mode $\cT_\m$. In other words, the 
latter has to be quite subdued, which is commensurable with the smallness of the parameter 
$\b$. In fact, it is easy to see that for $\b \ll 1$, the stable point $E_4$ located at the 
boundary (\ref{E-cp-exist}) of $\fM$ supports cosmologies which can be summed up as small 
deviations from $\L$CDM. A DE model of such sort has been the one studied earlier 
\cite{ssasb2}, in which statistical bounds on $\b$ are found by demanding that the value 
of, say, the Hubble constant $\Hp$ has to be within the $1 \s$ error limits of that 
predicted for $\L$CDM from physical observations. Although these bounds are important 
from the observational perspective, in order to see the overall qualitative nature of 
stable cosmologies represented by $E_4$ it suffices one to resort to the general 
(model-independent) upper limits --- the coarse one, viz. $\bmax = \rfra 3 2$, for the 
stable point $E_4$ to exist in the first place, and the tighter one, viz. $\bmax = \rfra 1 2$, 
in order that a phase of accelerated expansion of the universe is supported by $E_4$.

\subsection{Numerical solutions of the autonomous equations} \la{sec:numaut}

The limitation of the equilibrium solutions of the autonomous equations (or the critical 
points in the phase plane) is that they do not provide any quantitative information as to 
what the state of a system has been prior to reaching them. To overcome this (atleast 
partially), we require to find particular solutions of the autonomous equations. Doing so 
analytically is however a formidable proposition for the coupled set 
(\ref{E-auteq1})--(\ref{E-autcons}). We therefore resort to numerical techniques for a 
given range of initial values of the variables $X$ and $Y$. Each set of numerical 
solutions $X(N)$ and $Y(N)$ traces out a trajectory representing the system's evolution 
in the phase plane, right from the point of origin (in accord with the initial conditions) 
till the termination at one of the CPs. This also enables us to see the variation (with 
$N$) of any explicit function of $X$ or(and) $Y$, over the lifetime of a phase plane 
trajectory. Hence, in a cosmology resulting from the chosen initial conditions, we can 
in principle plot the quantities of interest, such as $\OX$, $\wx$ and $\sw$, over a 
significantly large span of time (or $N$). 
\begin{figure}[h]
\centering
\subfloat[{\footnotesize $\b=0.01$}]{\includegraphics[scale=0.572]{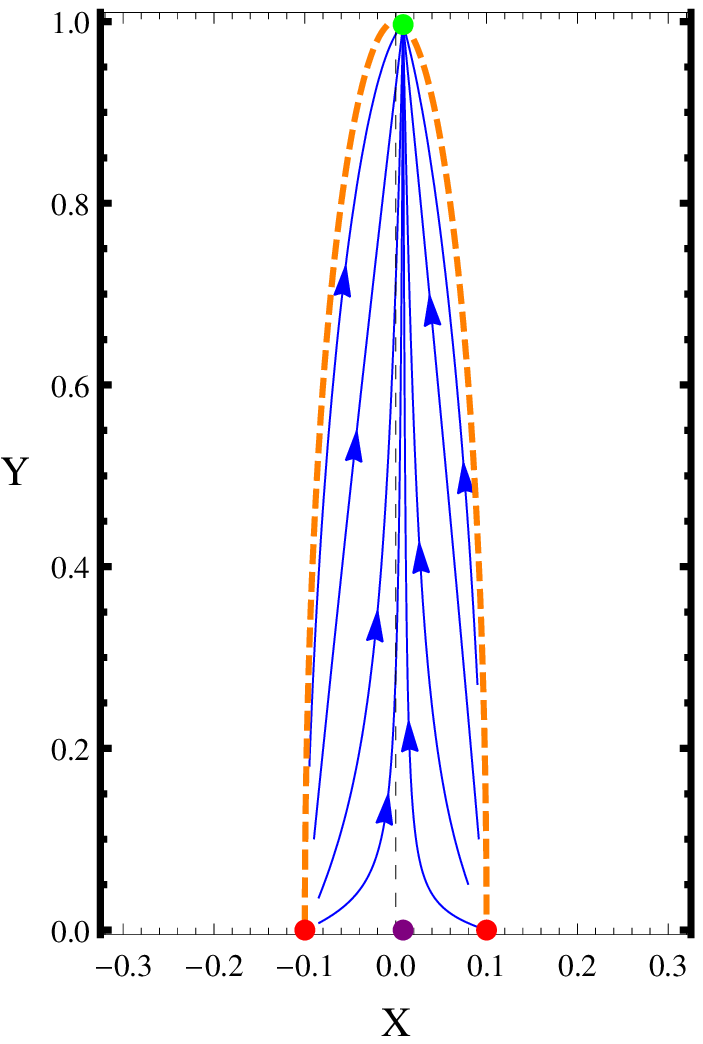}} ~
\subfloat[{\footnotesize $\b=0.05$}]{\includegraphics[scale=0.572]{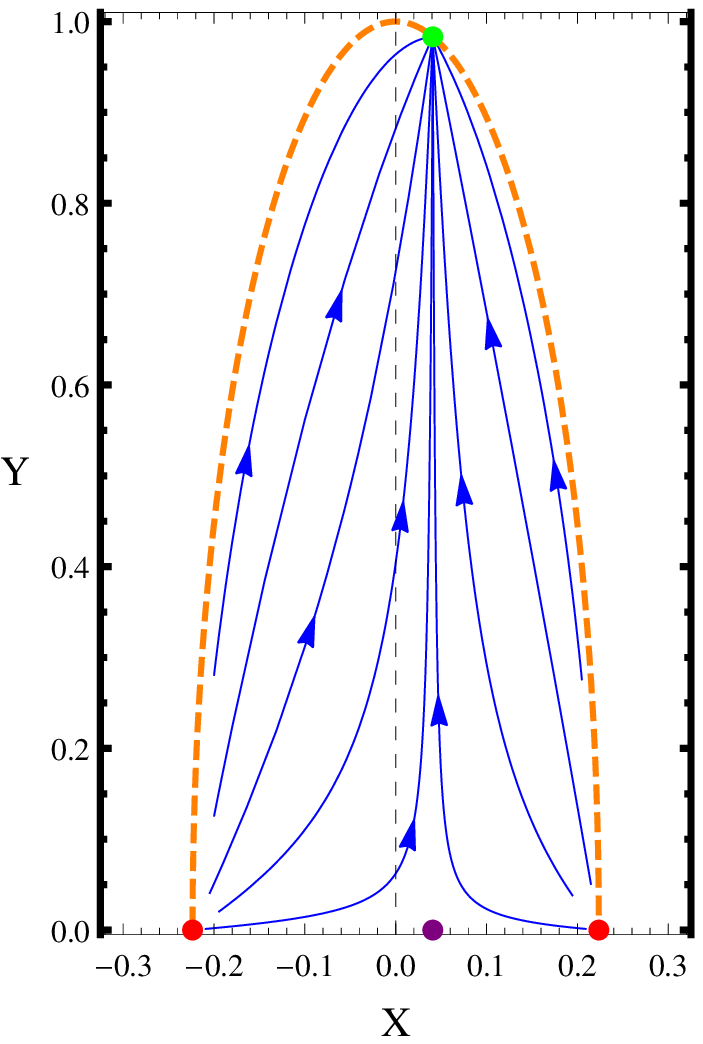}} ~
\subfloat[{\footnotesize $\b=0.1$}]{\includegraphics[scale=0.572]{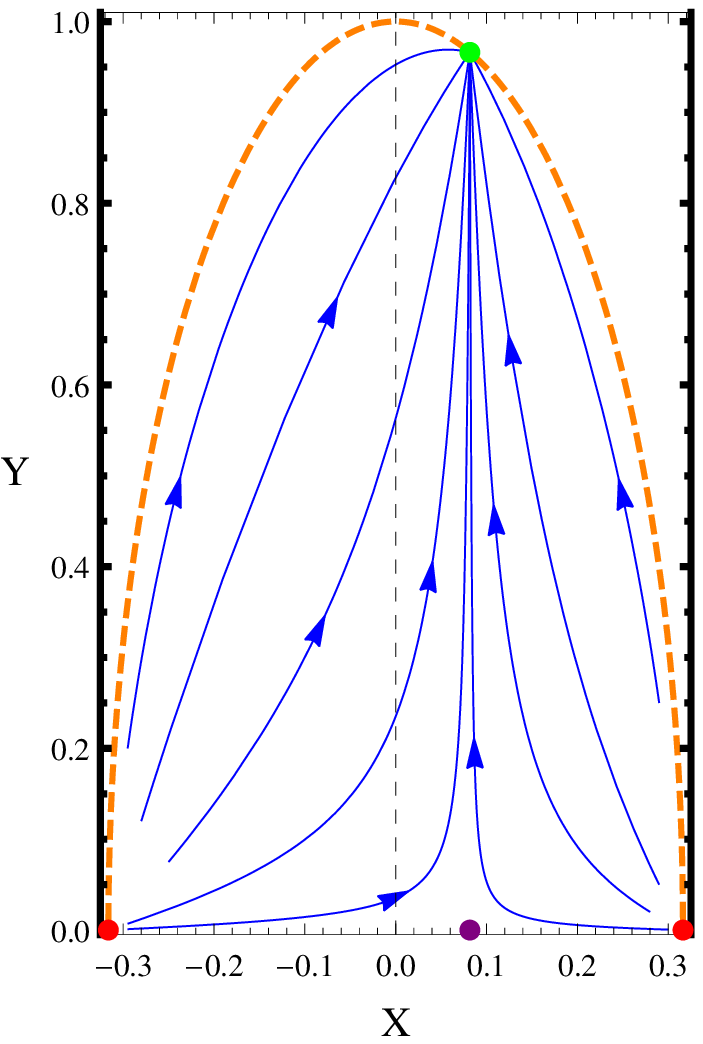}} \\
\subfloat[{\footnotesize $\b=0.5$}]{\includegraphics[scale=0.585]{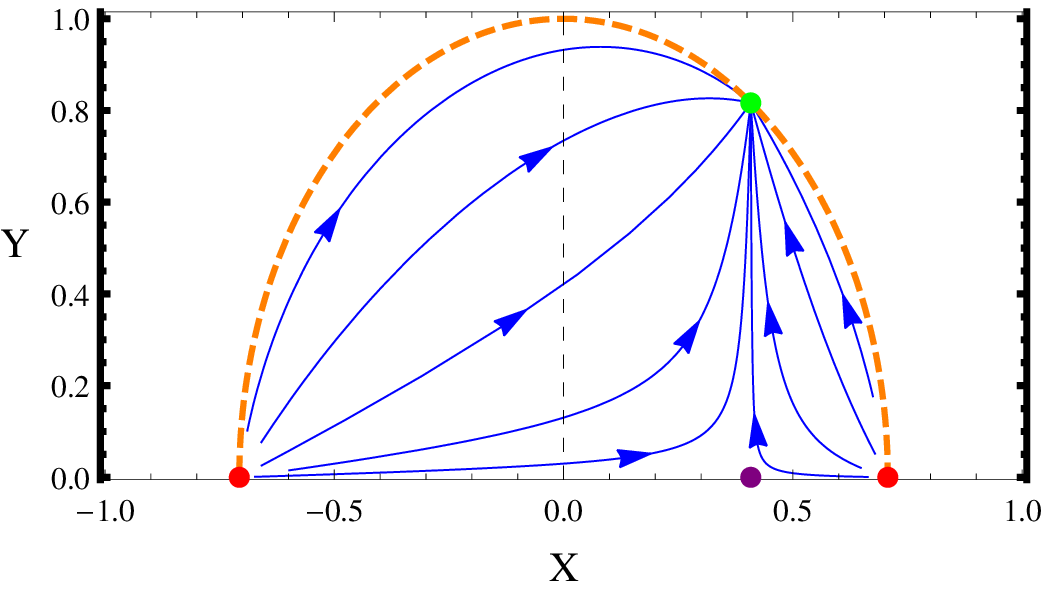}} ~
\subfloat[{\footnotesize $\b=1$}]{\includegraphics[scale=0.585]{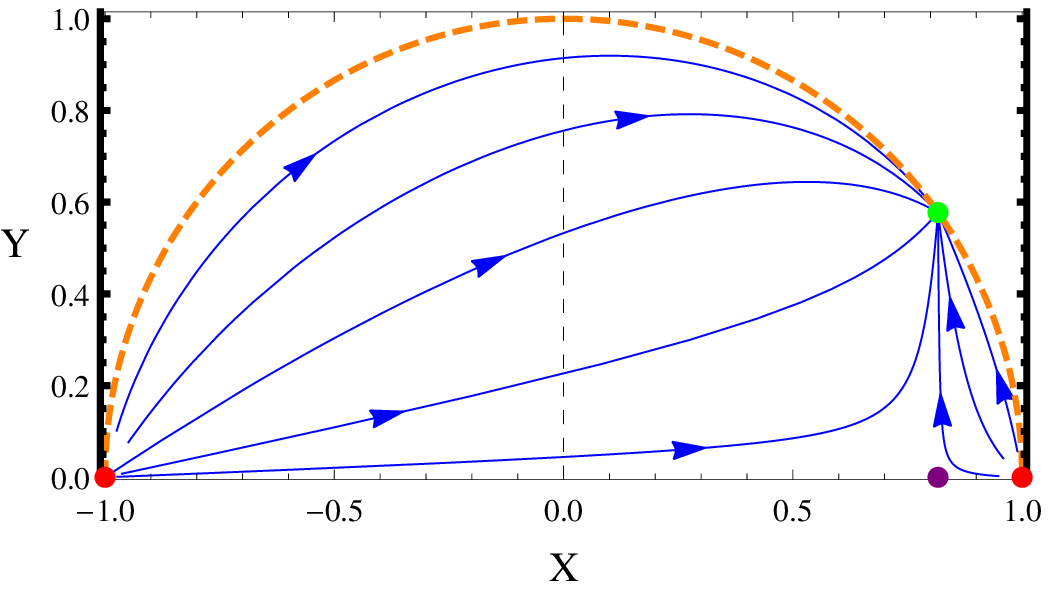}} 
\caption{\footnotesize Einstein frame phase portraits for different values of the parameter $\b$. The dots 
represent the critical points, arrows mark the direction of time-evolution along trajectories and the dashed 
curve demarcates the permissible region for cosmologies with non-negative effective matter density.}
\la{E-PPfig}
\end{figure}
Figs. \ref{E-PPfig}(a) -- (e) show the trajectories the system follows in the $XY$ plane to reach a CP
for different initial values and for the parametric settings $\b = 0.01, 0.05, 0.1, 0.5$ and $1$. As 
mentioned in the last subsection, there are actually four CPs to be taken into account, viz. those in 
the physically admissible region $\fM$, or on its elliptic boundary (\ref{E-cp-exist}). These CPs are 
superimposed for comparison in each of the Figs. \ref{E-PPfig}(a)--(e).

Our main interest however, from the point of view of the accelerating cosmologies, is in the Figs. 
\ref{E-PPfig}(a), (b) and (c), corresponding to the settings $\b = 0.01, 0.05$ and $0.1$ respectively. 
Figs. \ref{E-PPfig}(d) and (e), corresponding to $\b = 0.5$ and $1$ respectively, are only for the sake 
of completeness in the illustration. We see that the trajectories originating somewhere in the elliptically 
bounded region $\fM$ of the phase plane, excluding the $X$-axis, terminate at the stable point $E_4$. On 
the other hand, the trajectories which originate at a point on the $X$-axis, tend to terminate at the saddle 
point $E_3$. In fact, the trajectories originating anywhere in $\fM$ except the $X$-axis are deflected in 
a direction parallel to the $X$-axis towards $E_3$, which deflects them in a direction vertically above it, 
i.e. towards $E_4$ situated at $\le(X = \sq{\rfra 2 3} \b, \, Y = \sq{1 - \rfra{2\b} 3}\ri)$. In other 
words, the saddle point $E_3$ has the effect of {\em funnelling} trajectories towards the stable point $E_4$. 
Of course, the latter exhibits its own {\em attractive} nature as well. The $X$-axis (i.e. $Y=0$) is the 
stable axis for $E_3$ and the line $X = \sq{\rfra 2 3} \b$ is its unstable axis. As $\b$ is increased from 
a small value (say, $0.01$), the area of the region $\fM$ increases, with the decrease in the eccentricity 
of its elliptic boundary. Accordingly, the saddle point $E_3$ and the stable point $E_4$ shift away from 
the $Y$-axis, and so do the unstable points $E_1$ and $E_2$. The shift continues till $b \rightarrow \rfra
3 2$, whence the CPs $E_2, E_3$ and $E_4$ tend to coincide with the CP $E_5$ (not shown in the Figs. 
\ref{E-PPfig}(a) -- (e)), which approaches the elliptic boundary of the region $\fM$ from outside. Further 
increase in the value of $\b$ (beyond $\rfra 3 2$) would take $E_1, E_2$ and $E_3$ outside the physical 
realm, whereas $E_4$ and $E_5$ would cease to exist. 

Let us now examine the evolution of some cosmological parameters of interest, and the torsion parameters,
along a fiducial trajectory corresponding to a particular setting, say $\b = 0.01$. As to the initial 
conditions for this fiducial trajectory, we may conveniently set them at the present epoch ($N = 0$), i.e. 
by appropriately choosing the values of $X (0)$ and $Y (0)$. One obvious choice is that in line with the 
exact solution obtained in paper I, whose stability we have already established in 
this paper. Such a solution corresponds to taking the ansatz $\, e^\t = a^{6 \b}$, whence it follows from 
Eq. (\ref{E-aut}) that $X =$ constant $= \sq{\rfra 2 3} \b$. Moreover, since $\t (0) = 0$ (or $F (0) = 0$ 
by the definition (\ref{tauF})), we have from Eq. (\ref{E-effdens1}) $\, \Omp \equiv \Omf (0) = 1 - 
\rfraa{X^2 (0)}{\b} - Y^2 (0)$. Considering now the fiducial values:

\vskip 0.1in
\no 
(i) $\, \Omp = 0.3$ (which is roughly the observational prediction \cite{wmap,jla,planck} for most 
of the model-independent and model-dependent DE parametrizations), and 

\vskip 0.05in
\no 
(ii) $\, \b = 0.01$ (which is the order of magnitude estimation \cite{ssasb2} for the above ansatz, 
using the WMAP and Planck results \cite{wmap,planck}), 

\vskip 0.1in
\no 
we have the initial conditions
\be \la{E-ini}
X (0) = \sq{\fr 2 3} \b = 0.0082 \qquad \mbox{and} \qquad
Y (0) = \sq{1 - \fr{2 \b} 3 - \Omp} = 0.8327 \,\,.
\ee
\begin{figure}[h]
\centering
\subfloat[{\footnotesize Density parameters vs. $N$}]{\includegraphics[scale=0.62]{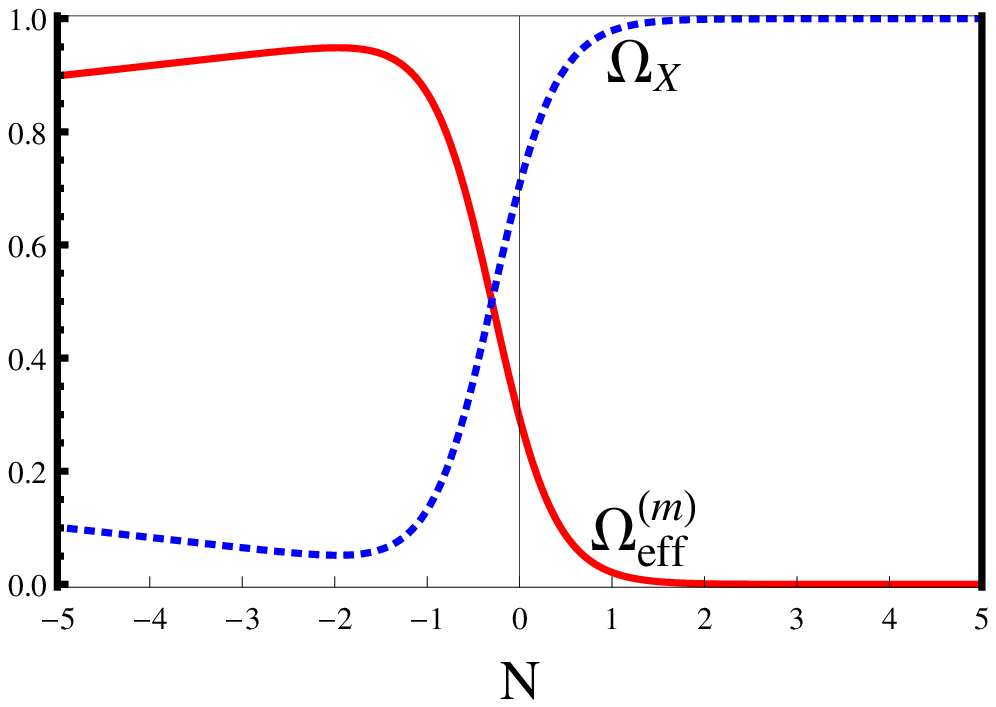}} ~
\subfloat[{\footnotesize EoS parameters vs. $N$}]{\includegraphics[scale=0.62]{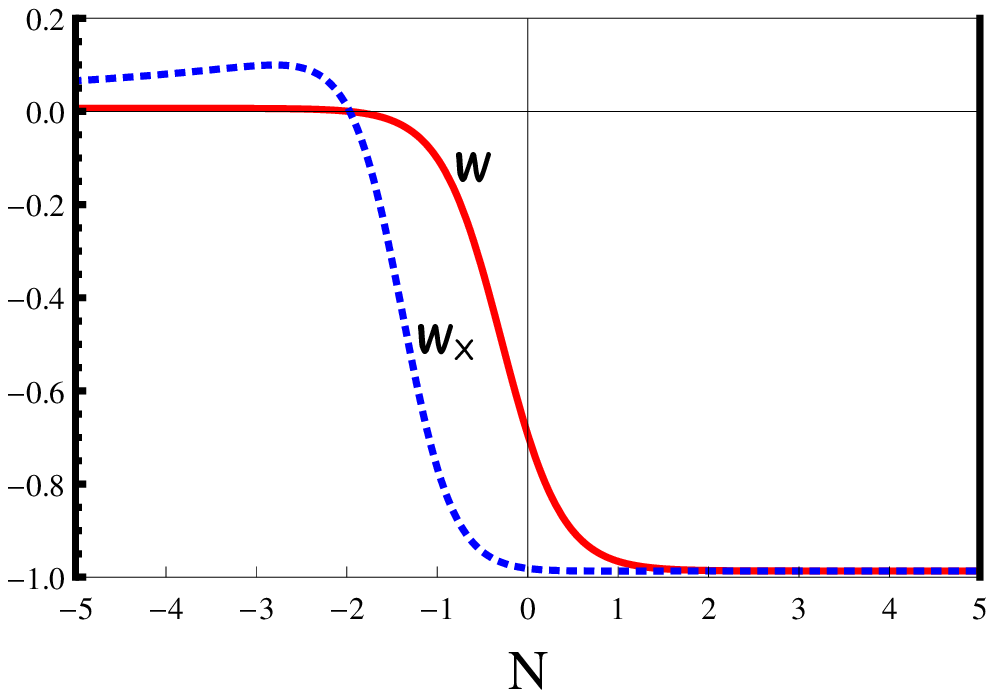}} 
\caption{\footnotesize Evolution of (a) the density parameters $\Omft$ and $\OXt$, and (b) EoS
parameters $\wx$ and $\sw$, over the trajectory with initial conditions $[X(0) = 0.0082, Y(0) = 
0.8327]$ for $\b = 0.01$, $\Omp = 0.3$.}
\la{E-PPF-para}
\end{figure}
Fig. \ref{E-PPF-para}(a) shows the evolution of the density parameters $\Omf (N)$ and $\OX (N)$ 
over the fiducial trajectory (with the above initial conditions) for a fairly wide range of 
e-foldings $N \in \{-5,5\}$\footnote{considering that $N = -5$ corresponds to a redshift $z 
= e^{-N} - 1 = 147.4$.}. As expected (in view of the small value of $\b$), these parameters 
vary with $N$ similar to their $\L$CDM analogues. There are some subtleties however. Note that 
the curves in Fig. \ref{E-PPF-para}(a) are asymmetric about $N = 0$. Actually, as we go from 
the present ($N = 0$) to the future regime ($N > 0$), $\Omf$ and $\OX$ rapidly approach a 
near-saturation to the values $0$ and $1$ respectively. That is, the DE tends to dominate
completely over the dust-like matter even in the not-so-distant future, which is quite 
identical to the case in $\L$CDM. On the other hand, as we go back in the past ($N < 0$), 
$\Omf$ and $\OX$ first approach each other rapidly, reach an equality point, then diverge 
with the same rapidity, attain extremum values, and finally approach each other once again 
(albeit very slowly) further back in the past. This is of course dissimilar to what happens 
for $\L$CDM, and its root cause can be traced to the original interaction between the torsion 
scalar $\t$ and the cosmological fluid. Although this interaction gets obscured in the 
effective picture, it leaves its imprint on the density profiles. In fact, the dissimilarity 
with $\L$CDM is also evident from the evolution patterns of the EoS parameters, viz. $\wx (N)$ 
and $\sw (N)$ corresponding to the dynamical DE and the system respectively, over the fiducial 
trajectory. These are shown in Fig. \ref{E-PPF-para}(b), for the same range of $N$, viz. 
$\{-5,5\}$.

As to the evolution of the torsion scalar $\t$, and that of the torsion parameters $|\cT|$ and $|\cA|$, 
over the fiducial trajectory, first note that since $X = \sq{\rfra 2 3} \b$ is constant throughout,
Eq. (\ref{tauF}) implies $F (N) = \sq{\rfra 2 3} \b N$. Therefore, $\t (N) = 6 \b N$, i.e. $\t$ varies 
linearly with $N$, as shown in Fig. \ref{E-PPF-tor}(a). Using Eqs. (\ref{E-tau}) and (\ref{E-aut}) we can 
now work out the functional forms of the parameters $|\cT|$ and $|\cA|$, and plot them after appropriate 
dimensional scaling. We choose to scale $|\cT|$ with the Hubble parameter $H$, in order to have a direct 
measure of the effect of the trace mode of torsion on the cosmological evolution. However, $|\cA|$ being 
a constant, it is imperative to consider its ratio with $|\cT|$ and see how that evolves with $N$, for 
the fiducial setting. The expressions of $\rfraa{\le|\cT\ri|} H$ and $\rfraa{\le|\cA\ri|}{\le|\cT\ri|}$ 
are found to be
\be \la{E-F-TA}
\fr{\le|\cT\ri|} H \,=\, 6 \b \, e^{2 \b N} \qquad \mbox{and} \qquad 
\fr{\le|\cA\ri|}{\le|\cT\ri|} \,=\, \fr{2 \, Y}{\b} \,\,.
\ee
\begin{figure}[h]
\centering
\subfloat[{\footnotesize Torsion scalar and trace 
parameter vs. $N$}]{\includegraphics[scale=0.62]{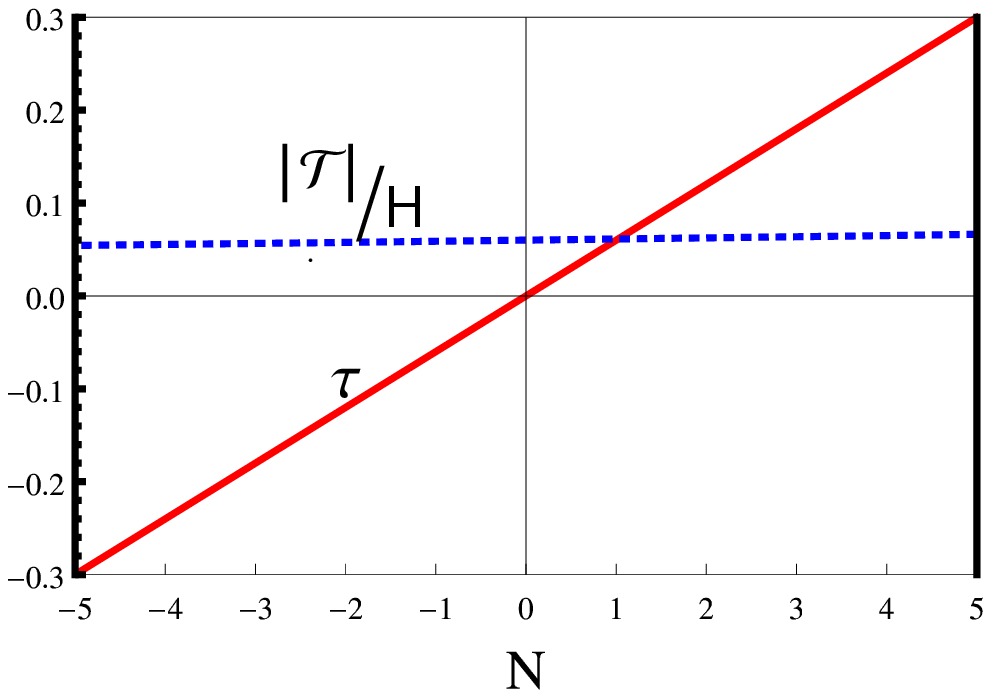}} ~
\subfloat[{\footnotesize Torsion pseudo-trace to 
trace ratio vs. $N$}]{\includegraphics[scale=0.62]{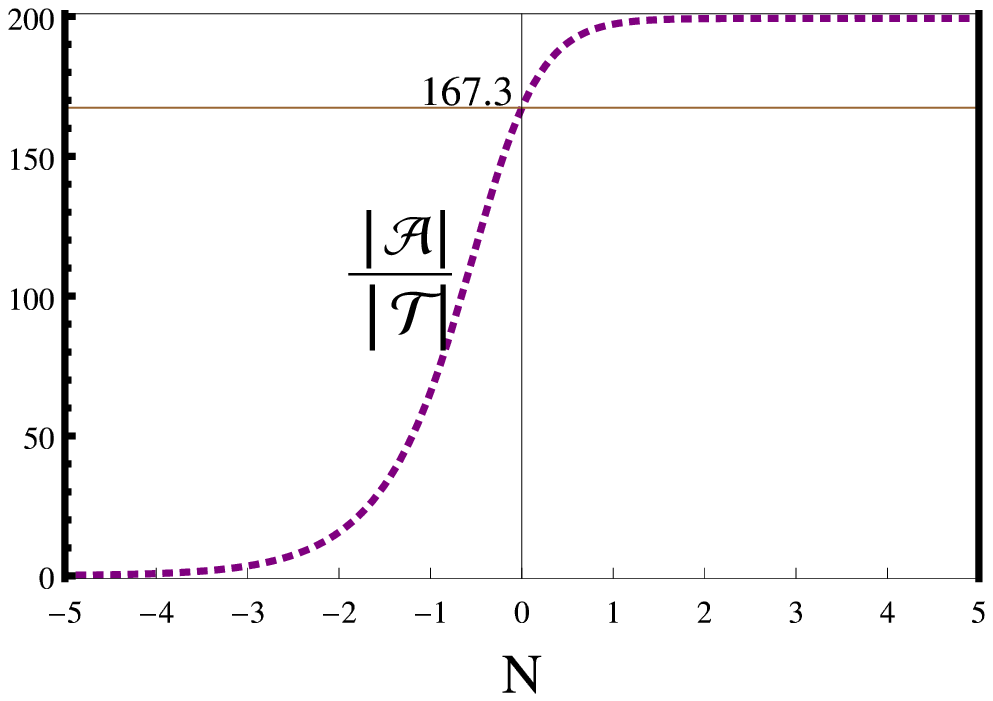}} 
\caption{\footnotesize Evolution of (a) the torsion scalar $\t$ and the trace parameter $|\cT|$ 
(in units of the Hubble parameter $H$), and (b) the ratio of the torsion pseudo-trace 
and trace parameters, $|\cA|$ and $|\cT|$, over the trajectory with initial conditions 
$[X(0) = 0.0082, Y(0) = 0.8327]$ for $\b = 0.01$, $\Omp = 0.3$.}
\la{E-PPF-tor}
\end{figure}
Fig. \ref{E-PPF-tor}(a) shows almost a constant variation of $\rfraa{\le|\cT\ri|} H$ with $N$, for the
entire range of $N \in \{-5,5\}$. This is expected, since the exponential factor in Eq. (\ref{E-F-TA}) 
hardly affects the evolution for the small value of $\b$ ($= 0.01$). Moreover, the magnitude of 
$\rfraa{\le|\cT\ri|} H$ is small as well ($\sim 6 \b = 0.06$), implying that the torsion trace has a
minor contribution to the Hubble rate throughout the evolution (in the above range of $N$). The 
evolution of the ratio $\rfraa{\le|\cA\ri|}{\le|\cT\ri|}$ over the fiducial trajectory is shown in
\ref{E-PPF-tor}(b). Deep in the past, this ratio has been very small, meaning a complete domination
of $\le|\cT\ri|$ over $\le|\cA\ri|$. However, as $\le|\cT\ri|$ decreases with $N$, the ratio grows
and in the near past the growth becomes very rapid. At the present epoch its value is $\rfraa{2 Y(0)}
\b = 167.3$. It continues to increase in the future until saturating to a value of about $200$.

Overall thus, in the MST scenario we have a DE evolution slightly deviated from $\L$CDM, when the 
norm of the pseudo-trace of torsion is about two orders of magnitude stronger than that of its
trace, in the near past and future.

\section{Phase plane analysis in the Jordan frame} \label{sec:J-PP}

Recall that while dealing with the MST formalism in a standard cosmological setup in section 
\ref{sec:mst}, we first arrived at the scalar-tensor action (\ref{J-ac}) in the Jordan frame 
with a characteristic non-minimal coupling between the torsion scalar $\t$ and gravity. We 
then went on to obtain the corresponding Einstein frame action (\ref{E-ac}) and carried out 
the phase plane analysis in the Einstein frame, because it is straightforward and in line 
precisely with the canonical formulation of GR. Having performed the stability analysis of the 
Einstein frame solutions in section \ref{sec:E-PP}, we now turn our attention to their Jordan 
frame counterparts in this section. 

It should be pointed out here that the Jordan frame action (\ref{J-ac}) can always be cast
in the standard Brans-Dicke (BD) form \cite{frni,fujii,ssasb2}, once we identify the BD
scalar field to be $\keff^{-2} (\t) = \k^{-2} e^{\rfra{2 \t \!} 3}$ and the effective BD parameter 
to be
\be \la{BD-param}
\fw \,=\, \fr{1 - 6 \b}{4 \b} \,=\, \fr{9 \k^2 \vep} 4 \,\,,
\ee
where $\vep$ is the dimensionful parameter [{\it cf}. Eq. (\ref{Ldef})], that appears in the 
kinetic term of the torsion scalar $\t$ in Eq. (\ref{J-ac}). Throughout the analysis in this
section, we shall categorize the results in terms of $\fw$, instead of $\b$ or $\vep$. The 
advantage is that the analysis could be restricted by the lower bound on $\fw$ ($\simeq 40$, to 
as large as $40000$) estimated from a plethora of independent studies \cite{BDbounds1,BDbounds2}. 
Of course, we also have the theoretical limitation $\fw > - \rfra 3 2$, in accord with the 
presumption $\b > 0$.

\subsection{Cosmological equations and the effective scenario} \la{sec:J-eqs}

For the spatially flat FRW metric $\, g_{\m\n} =$ diag$[-1, \, a(t), \, a(t), \, a(t)]$, 
where $t$ is the Jordan frame comoving time coordinate, we simply have $\t = \int dt 
\le|\cT\ri|$. The corresponding action (\ref{J-ac}) leads to Friedmann and Raychaudhuri 
equations:
\be \la{J-eq} 
H^2  = \fr{\keff^2} 3 \, \rJ \qquad \mbox{and} \qquad
\dot{H} = - \, \fr{\keff^2} 2 \le[\rJ + \pJ\ri] \,,
\ee
where $\, H := \rfrac{\dot{a}} a$ is the Jordan frame Hubble parameter, and 
\bea 
&& \rJ = \rmt + e^{\rfra{2 \t \!} 3} \le[\fr{2 \fw \dot{\t}^2}{9 \k^2} + \L - 
\fr{3 H \dot{\t}}{\k^2}\ri] \,,
\la{J-dens} \\
&& \pJ = e^{\rfra{2 \t \!} 3} \le[\fr{2 \ddot{\t}}{3 \k^2} + \fr{2 (2 + \fw) \dot{\t}^2}
{9 \k^2} - \L + \fr{4 H \dot{\t}}{3 \k^2}\ri] \,,
\la{J-pres}
\eea
are the corresponding total (or critical) energy density and pressure, $\, \rmt$ being the 
energy density of the cosmological matter in the form of the pressureless dust. Unlike the
situation in the Einstein frame, the dust density $\rmt$ is conserved here, implying 
$\, \rmt = \rmp a^{-3}$, where $\, \rmp = \rmt \rvert_{a=1}$ is the present-day value of 
$\rmt$. That is, the `dust' has its usual interpretation in the Jordan frame. However, the
critical density $\rJ$ is itself satisfies a somewhat non-standard conservation relation:
\be \la{J-consv}
\dot{\r}_{\!_J} \,+\, 3 \, H \le[\rJ + \pJ\ri] =\, \fr{2 \, \dot{\t}} 3 \, \rJ \,\,.
\ee
Therefore, while making predictions for the Jordan frame observables, one requires to take 
account of the rate at which the torsion scalar $\t$ changes (or, the gravitational coupling 
varies\footnote{Since the coupling factor $\keff (\t) = \k \, e^{\rfra{- \t \!} 3}$, we have 
$\, \rfraa{\dot{\t}} 3 = - \, \rfraa d {dt} \le[\ln \keff\ri]$.}) with time.  This is 
reminiscent of the weirdness of all Brans-Dicke type of cosmological models, especially 
when it comes to making statistical estimates of the changes due to such models on the 
values of cosmological parameters predicted by some known model, for e.g. $\L$CDM. One way 
to avoid this, i.e. bypass the direct confrontation with the running ($\t$-dependent) 
gravitational coupling, is to define and decompose the critical density similar to that
in a minimally coupled theory \cite{ssasb2}: 
\be \la{J-criteff}
\r \,:=\, \fr{3 \, H^2}{\k^2} \,=\, \rJ \, e^{\rfra{- 2 \t \!} 3} =\, \rmt \,+\, \rx \,\,,
\ee 
where $\, \rmt = \rmp a^{-3}$ is the dust density, $\rx$ is a surplus density, which we
shall consider as the DE density. We demand $\r$ to satisfy the usual conservation relation:
\be \la{J-consveff}
\dot{\r} \,+\, 3 \, H \le[\r + p\ri] =\, 0 \,\,,
\ee
where $p$ is the total effective pressure. Under such a demand, we have the effective DE 
density and pressure, given respectively as
\be \la{J-de-denspr}
\rx = \rJ \, e^{\rfra{- 2 \t \!} 3} - \rmt \qquad \mbox{and} \qquad
\px = p = \pJ \, e^{\rfra{- 2 \t \!} 3} \,\,,
\ee
satisfying the conservation relation
\be \la{J-de-consv}
\dot{\r}_{\!_X} \,+\, 3 \, H \le(\rx + \px\ri) \,=\, 0 \,\,.
\ee
It should however be pointed out here that the phase plane analysis in the Jordan frame can in 
principle be carried out with either of the sets $\le\{\rJ, \pJ, \rmt\ri\}$ and $\le\{\r, \rmt, 
\rx, \px\ri\}$. The stability criterion for the system, determined from the autonomous equations 
(to be constructed), is anyway not commensurate with the choice of the system constituents. Since
the analysis for the original description of the system in terms of the critical density $\rJ$ 
[{\it cf}. Eq. (\ref{J-dens})] would, in many respects, be similar to that for the standard 
scalar-tensor cosmologies in the Jordan frame \cite{tsuj}, we shall proceed to carry out that 
first. Then in due course, we shall examine the characteristic changes in the effective scenario, 
described by the Eqs. (\ref{J-criteff})--(\ref{J-de-consv}), most importantly with the objective 
of finding a stable DE model.

\subsection{Autonomous equations and the physically admissible region(s)} \la{sec:J-aut}

Defining the phase space variables as:
\be \la{J-aut}
X \,:=\, \fr{2 \, \dot{\t}} {3 \sq{6} H} \qquad \mbox{and} \qquad
Y \,:=\, \fr \k H \sq{\fr{\L} 3} \,\,\,,
\ee
we obtain from Eqs. (\ref{J-eq})--(\ref{J-consv}) the autonomous equations
\bea
&& \fr{dX}{dN} \,=\, \fr 3 {2 \fw + 3} \le(\fw X^2 - Y^2 - \sq{6} X - 1\ri)
\le[\le(\fw + 1\ri) X - \fr 1 {\sq{6}}\ri] \,, 
\la{J-auteq1} \\
&& \fr{dY}{dN} \,=\, \fr{3 \, Y}{2 \fw + 3} \le[\le(\fw + 1\ri) \le(\fw X^2 - Y^2 + 1\ri)
- \sq{\fr 2 3} \fw X + 1\ri] \,. \la{J-auteq2}
\eea
Moreover, the conventional forms (\ref{J-dens}) and (\ref{J-pres}) of the Jordan frame 
critical density $\rJ$ and the total pressure $\pJ$, lead to the following expressions 
for the corresponding matter density parameter $\Om_J$ and the total EoS parameter $\wJ$:
\bea 
&& \Om_J := \fr{\rmt}{\rJ} \,=\, 1 \,-\, \fw \, X^2 \,-\, Y^2 \,+\, \sq{6} \, X \,\,, 
\la{J-m-dens} \\
&& \wJ := \fr{\pJ}{\rJ} \,=\,
\fr 2 {2 \fw + 3} \le[\le(\fw + 1\ri) \le(\fw \, X^2 \,-\, Y^2\ri) -\, 
\sq{\fr 2 3} \, \fw X \,+\, \fr 1 2\ri] \,.
\la{J-w}
\eea
Similar to what we have seen in the Einstein frame, the autonomous system is symmetric under 
$Y \rightarrow -Y$. All points in the $XY$ phase plane do not represent physically relevant 
solutions as well. However, as opposed to Eq. (\ref{E-effdens1}) for the Einstein frame matter 
density parameter $\Omf$, the above Eq. (\ref{J-m-dens}) for the corresponding parameter
$\Om_J$ in the Jordan frame does not mean that that matter density would keep on evolving 
with time (or $N$) even after the system reaches a critical point (CP). Consequently, in order
to comply with the physical relevance condition $0 \leq \Om_J \leq 1$ in the Jordan frame, 
a CP is not required to lie on an ellipse that corresponds to vanishing matter density, as 
in the Einstein frame. Instead, for all points (critical or otherwise) in the phase plane, Eq. 
(\ref{J-m-dens}) defines the {\em physically admissible} region(s) $\fM$, bounded from both 
inside and outside by the curves
\be \la{J-physrel}
\fw X^2 +\, Y^2 -\, \sq{6} \, X \,=\, C \,\,, \qquad \mbox{where} \quad C = 0 ~\mbox{and}
~ 1 \,, ~ \mbox{respectively}\,.
\ee 
These curves evidently have identical shape, typified by conic sections of all possible sorts
(viz. circle, ellipse, parabola and hyperbola)\footnote{Note that, when the curves are open 
(i.e. parabolic or hyperbolic), they are actually the `demarcation lines' in the phase plane, 
rather than `boundaries'. Moreover, for hyperbolic demarcations the sets of disjoint branches 
would define two disjoint regions $\fM$, instead of a solitary one otherwise.}, depending on
the value of the parameter $\fw$, as examined below:  

\vskip 0.1in
\no 
{\bf Case I}. $\, \fw > 0$ : Eq. (\ref{J-physrel}) can be recast as
\be \la{J-phys-ell}
\fr{\le(X - \s\ri)^2}{b_x^2} + \fr{Y^2}{b_y^2} = 1 \,\,, \qquad \mbox{where} \quad 
\s = \fr{\sq{3}}{\sq{2} \fw} \,\,, \quad b_y^2 = \fw \, b_x^2 = \fr 3 {2 \fw} + C \,\,.
\ee
So the curves corresponding to $C = 0$ and $1$ in general represent concentric ellipses,
an inner and an outer one respectively, which enclose a solitary region $\fM$ accommodating 
the physically relevant trajectories in the phase plane. This region is symmetric about 
about the point $(\s,0)$. Moreover, the inner ellipse passes through the origin, i.e. one 
of its antipodes (or vertices), along either the minor axis or the major axis, is coincident 
with the origin. Consider further the following sub-cases:

\vskip 0.1in
\no 
{\bf (i)} $\, \fw > 1$ : We have $\, b_x^2 < b_y^2$. So, $b_x$ and $b_y$ are respectively the 
semi-minor and semi-major axes of the elliptic boundaries. 
\begin{figure}[h]
\centering
\subfloat[{\footnotesize $\fw = 250$}]{\includegraphics[scale=0.55]{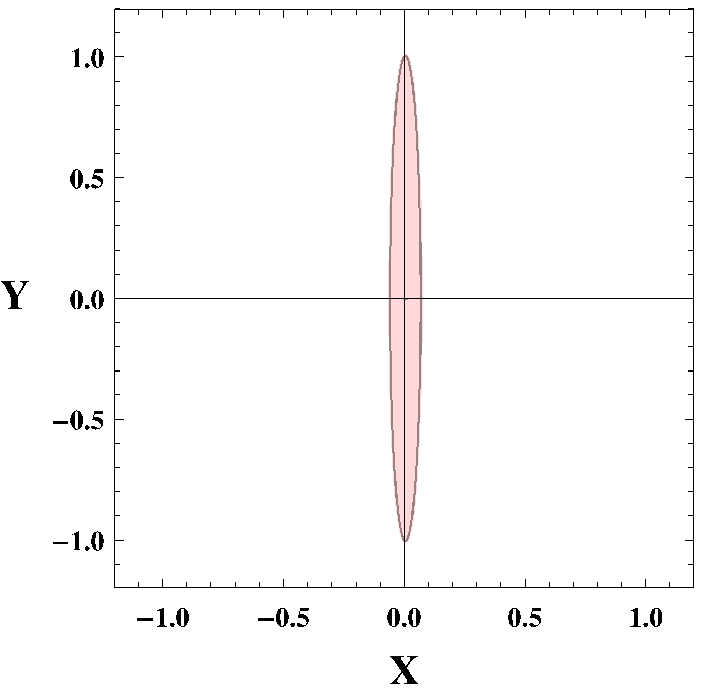}}\quad
\subfloat[{\footnotesize $\fw = 25$}]{\includegraphics[scale=0.55]{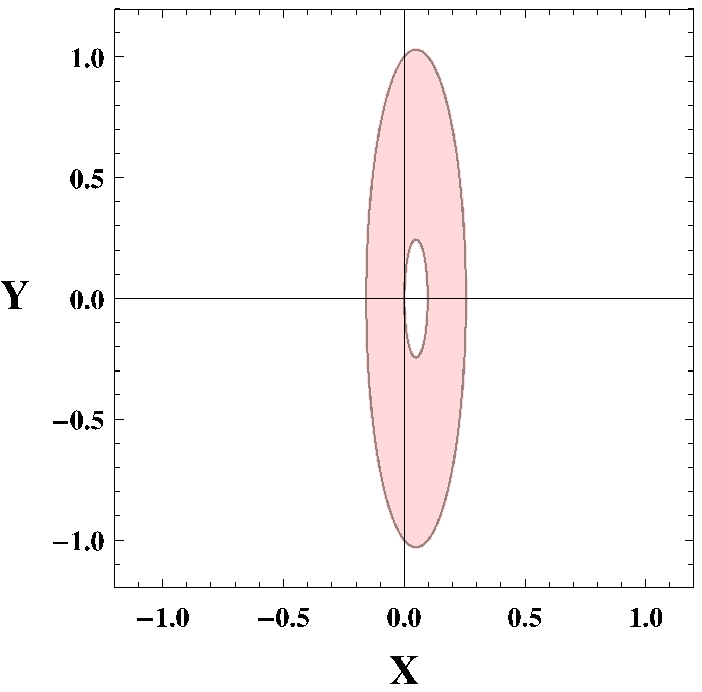}}\quad
\subfloat[{\footnotesize $\fw = 5$}]{\includegraphics[scale=0.55]{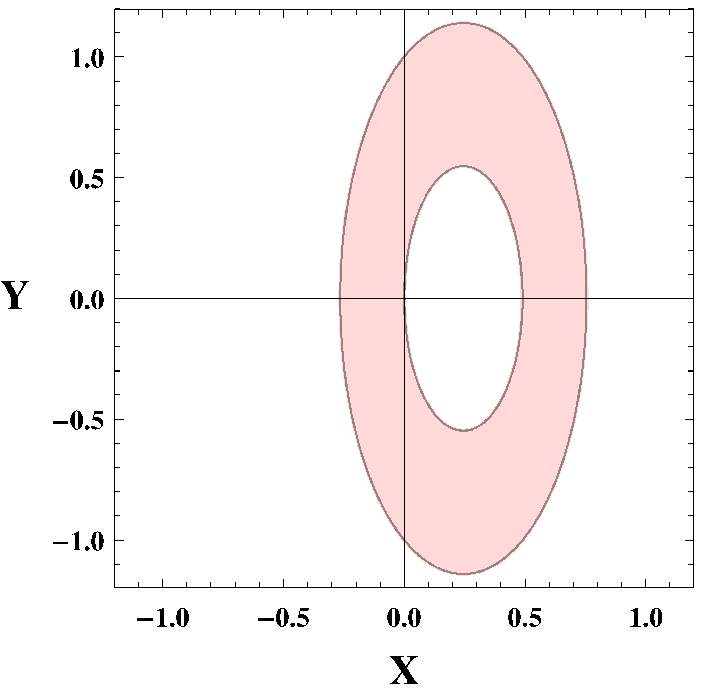}}\\
\subfloat[{\footnotesize $\fw = 1$}]{\includegraphics[scale=0.55]{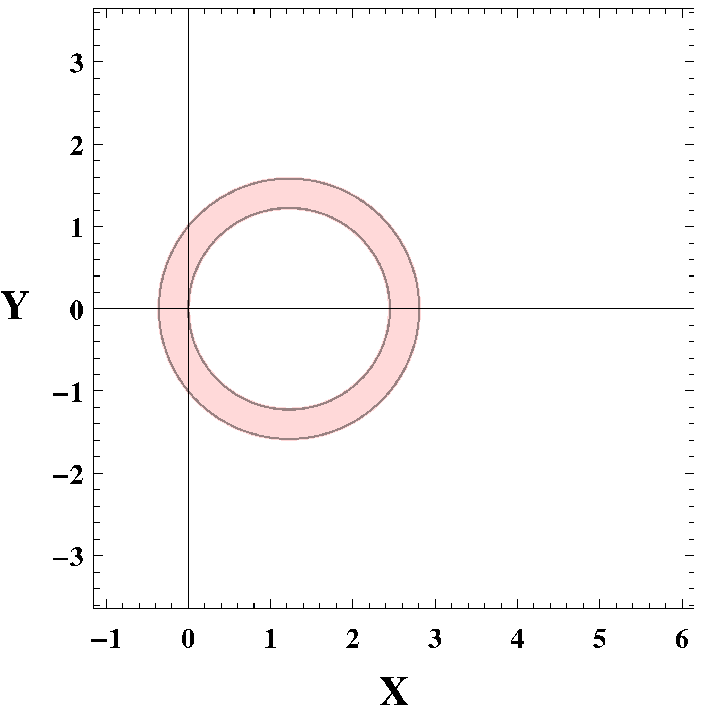}}\quad
\subfloat[{\footnotesize $\fw = 1$}]{\includegraphics[scale=0.55]{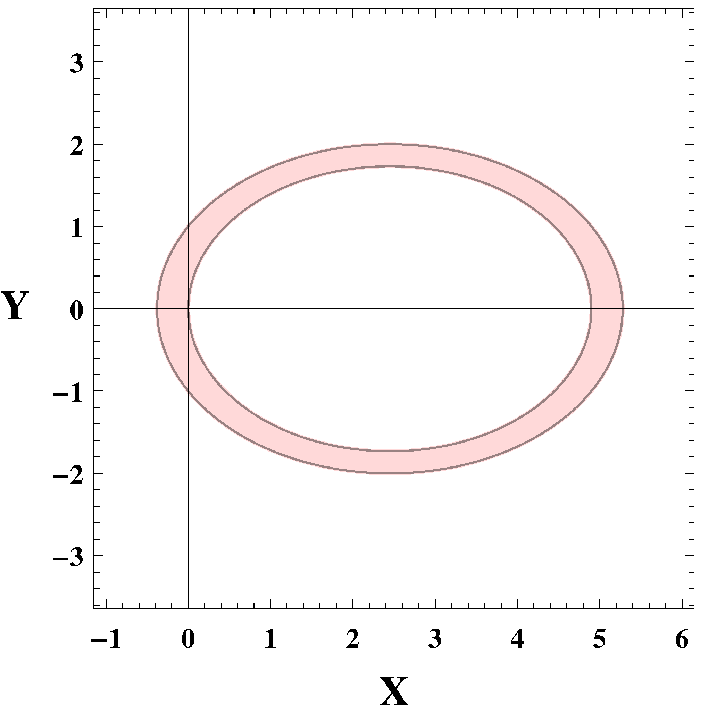}}\quad
\subfloat[{\footnotesize $\fw = 1$}]{\includegraphics[scale=0.55]{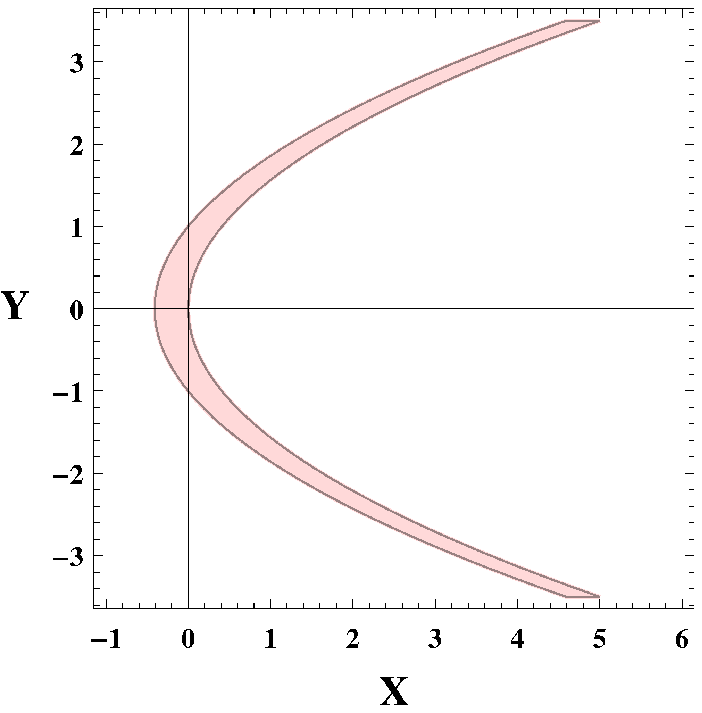}}
\caption{\footnotesize The physically admissible regions of the phase plane, marked by shades,
for different values of $\fw \geq 0$. Whereas the plots in the top row correspond to $\fw > 1$, 
those in the bottom row are for $0 \leq \fw \leq 1$. For a clear comparison of the plots in every 
row, their $X$-range and $Y$-range, as well as their axes calibration, are kept the same. }
\la{J-PPR-1}
\end{figure}
These ellipses have the same eccentricity $e = \sq{1 - \rfraa{b_x^2}{b_y^2}} = \sq{1 - \rfraa 
1 \fw}$, but differ in the focal length $\, f = e \, b_y$ and the semi-latus rectum $\, \ell =
\le(1 - e^2\ri) b_y$. Moreover, one vertex along the minor axis of the inner ellipse coincides
with the origin. Refer to Figs. \ref{J-PPR-1}(a), (b) and (c) for exemplary settings $\fw = 250$, 
$25$ and $5$ respectively, and with the same $X$-range, $Y$-range and axes calibration. As $\fw$ 
increases (from the unit value), we see the following:
\bit
\item The ellipses becomes more and more eccentric and when $\fw \rightarrow \infty$, $\, e 
\rightarrow 1$. 
\item Both the ellipses shift towards the origin along the $X$-axis, and their (common) 
center located at $(\s,0)$, coincides with the origin in the limit $\fw \rightarrow \infty$.
\item The semi-latus recta of the ellipses decrease, and for large $\fw$, $\, \ell 
\rvert_{_{C = 0}} \sim \fw^{- \rfra 3 2}$ whereas $\ell \rvert_{_{C = 1}} \sim \fw^{- 1}$, 
i.e. the reduction is faster for the inner ellipse compared to the outer one. The semi-major 
and semi-minor axes of the ellipses also decrease, and in the limit $\fw \rightarrow 
\infty$, $\, b_y\rvert_{_{C = 1}} \rightarrow 1$ whereas $b_y\rvert_{_{C = 0}} \rightarrow 0$. 
\eit
Hence, for a very large $\fw$ the physically admissible region $\fM$ nearly becomes a very 
eccentric elliptical zone about the origin, since its inner boundary gets almost obliterated
(see Fig. \ref{J-PPR-1}(a) as an example). Some weirdness is there though in the limit $\fw 
\rightarrow \infty$, since $e \rightarrow 1$, but none of the elliptic boundaries reduce to 
parabola. Instead, the region $\fM$ tends to become a line segment of unit length, coincident 
with the $Y$-axis and symmetric about the origin. Such a line segment in the phase plane is 
reminiscent of the $\L$CDM model, which one expects while taking the parametric limit $\b 
\rightarrow 0$ (or $\fw \rightarrow \infty$) in the MST-cosmological setup. In the other limit 
$\fw \rightarrow 1$, there is nothing unusual --- $e \rightarrow 0$ and the ellipses reduce 
to circles. 

\vskip 0.1in
\no 
{\bf (ii)} $\, \fw = 1$: We have $\, b_x^2 = b_y^2 = \rfra 3 2 + C$, and therefore both the 
boundaries of $\fM$ are concentric circles, with the radii $\sq{\rfra 3 2}$ and $\sq{\rfra 5 2}$
respectively for the inner one (passing through the origin) and the outer one. The common center 
is at $\big(\sq{\rfra 3 2}, 0\big)$, and the (annular) region $\fM$ has width $\sq{\rfra 5 2} - 
\sq{\rfra 3 2} = 0.3564$. See Fig. \ref{J-PPR-1}(d). 

\vskip 0.1in
\no 
{\bf (iii)} $\, 0 < \fw < 1$ : We have $\, b_x^2 > b_y^2$. So, $b_x$ and $b_y$ are respectively the 
semi-major and semi-minor axes for either of the bounding ellipses, having the same eccentricity
$e = \sq{1 - \rfraa{b_y^2}{b_x^2}} = \sq{1 - \fw}$, but different focal lengths $\, f = e \, b_x$ 
and semi-latus recta $\, \ell = \le(1 - e^2\ri) b_x$. Compared to the sub-case (i) above, the minor 
and major axes of the ellipses are interchanged (see Fig. \ref{J-PPR-1}(e) for the exemplary setting 
$\fw = 0.5$). As $\fw$ decreases (from the unit value), the eccentricity $e$ decreases, and the 
common center of the ellipses shifts more and more away from the origin along the $X$-axis. So the
region $\fM$ becomes more and more narrow (wide) along the common minor (major) axis. In the limit 
$\fw \rightarrow 0$, $\fM$ becomes a parabolic zone, discussed below.

\vskip 0.1in
\no 
{\bf Case II}. $\, \fw = 0$ : Eq. (\ref{J-physrel}) reduces to the equation of a parabola
\be \la{J-phys-par}
Y^2 \,=\, 4 \, f \le(X \,+\, \s\ri) \,\,, \qquad \mbox{with} \qquad f = \sq{\fr 3 8} \, 
\quad \mbox{and} \quad \s = \fr C {\sq{6}} \,\,.
\ee
The inner parabolic demarcation line has its vertex coincident with the origin, whereas the 
outer one has vertex shifted from the origin by an amount $\, \s = \rfra 1 {\!\!\sq{6}}$ along 
the negative $X$-axis. Both the parabolae have the same focal length $f = \sq{\rfra 3 8}$, and 
the region $\fM$ between them is a solitary zone that extends (with gradually diminishing width) 
to infinity along the $X$-direction (see Fig. \ref{J-PPR-1}(f)). Such a region in the phase 
plane is in fact quite unique since it depicts the bizarre scenario in which there is no kinetic 
term for the torsion scalar $\t$ in the Jordan frame action (\ref{J-ac}), and yet $\t$ is 
dynamical by virtue of its non-minimal coupling with gravity. 

\vskip 0.1in
\no 
{\bf Case III}. $\, \fw < 0$ : Eq. (\ref{J-physrel}) can now be recast as
\be \la{J-phys-hyp}
\fr{\le(X + \s\ri)^2}{b_x^2} - \fr{Y^2}{b_y^2} = 1 \,\,, \qquad \mbox{where}
\quad \s = \fr{\sq{3}}{\sq{2} |\fw|} \,\,,  \quad  
b_y^2 = |\fw| \, b_x^2 = \fr 3 {2 |\fw|} - C \,\,,
\ee
provided $\, |\fw| < \rfra 3 {2 C}$, which of course applies here for ($C = 0, 1$) by the 
presumption $\b > 0$. The curves represented by Eq. (\ref{J-phys-hyp}) are hyperbolae with 
major axis along the $X$-direction and the (common) center, i.e. the point of intersection 
of the asymptotes, at $(- \s, 0)$. Trajectories of physical relevance are therefore contained 
in two disjoint open regions $\fM$, between the pairs of hyperbolic branches (corresponding 
to $C = 0, 1$), which are symmetric about $(- \s, 0)$. The region on the right has the vertex 
of the inner hyperbolic demarcation line located at the origin, whereas the one on the left 
is entirely in the negative $X$ quadrant.  
\begin{figure}[h]
\centering
\subfloat[{\footnotesize $\fw = - 0.5$}]{\includegraphics[scale=0.55]{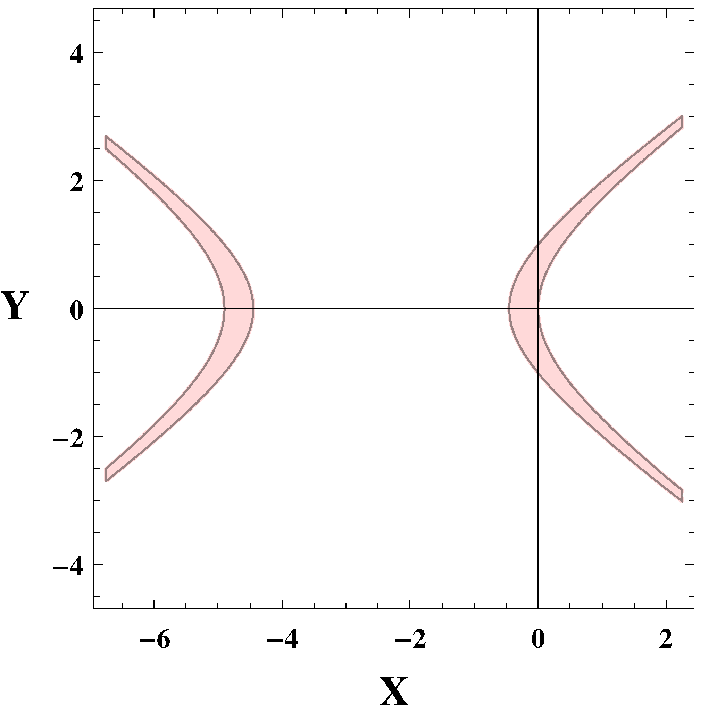}}\quad 
\subfloat[{\footnotesize $\fw = - 1$}]{\includegraphics[scale=0.55]{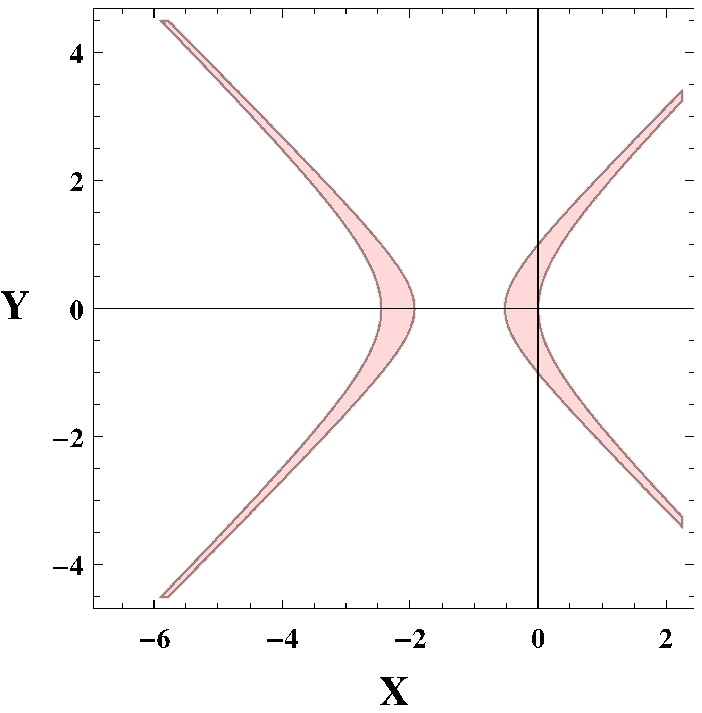}}\quad
\subfloat[{\footnotesize $\fw = - 1.5$}]{\includegraphics[scale=0.55]{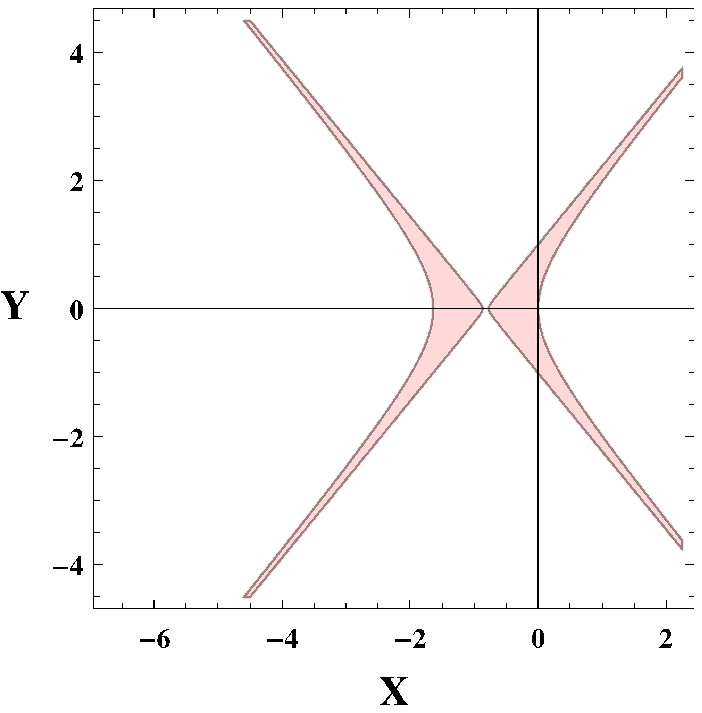}}
\caption{\footnotesize The physically admissible regions of the phase plane, marked by shades,
for different values of $\fw < 0$. The demarcation lines for these regions are hyperbolae, and
for the special setting $\fw = 1$, rectangular hyperbolae. In order to make a clear comparison of 
the plots, their $X$-range and $Y$-range, as well as their axes calibration, are kept the same.} 
\la{J-PPR-2}
\end{figure}
The inner and outer hyperbolae have the same eccentricity $e = \sq{1 + \rfraa{b_y^2}{b_x^2}} = 
\sq{1 + |\fw|}$, but different focal lengths $\, f = e \, b_x$ and semi-latus recta $\, \ell = 
\le(e^2 - 1\ri) b_x$. However, unlike the case I above, change in the values of $\fw$ do not 
make any major alteration in the characteristics of the curves, for e.g. their eccentricity, 
focal length, etc. The only point to note is that the curves become rectangular hyperbolae for 
$\fw = - 1$. Figs. \ref{J-PPR-2}(a), (b) and (c) illustrate these curves and the regions they 
demarcate, for exemplary settings $\fw = - 0.5$, $\fw = - 1$ and $\fw \approx - 1.5$ respectively. 
With the increase in $|\fw|$, the hyperbolae become more and more eccentric and their branches 
come closer together, i.e. their focal lengths decrease. Also, the (common) center of the 
hyperbolae move towards the origin along the $X$-axis. As $\, \fw \rightarrow - \rfra 3 2$, the 
two hyperbolic regions $\fM$ tend to merge together. In fact, $\, \fw \rightarrow - \rfra 3 2$ 
is a discontinuous limit, since the axes of hyperbolae get interchanged for $\fw \gtrless - 
\rfra 3 2$ (not shown in Fig. \ref{J-PPR-2}). Anyway, we exclude here the possibility $\fw 
< - \rfra 3 2$, because it implies $\b < 0$.

\subsection{Critical points and the dynamical evolution of the universe} \la{sec:J-dyn}

Following the procedure described in subsection \ref{sec:E-aut}, we find that corresponding to 
the Jordan frame autonomous Eqs. (\ref{J-auteq1}) and (\ref{J-auteq2}) there could be {\it five} 
distinct critical points (CPs). These CPs $(X_c, Y_c)$ are listed in Table \ref{J-tab-roots}, 
alongwith the domains of their physical relevance, given by the appropriate range of values of 
$\fw$.
\begin{table}[phtb]
\tbl{Jordan frame critical points and the parametric domains of their physical relevance.}{
\begin{tabular}{@{}ccc@{}} \toprule
CP: & ~ Location $\le(X_c, Y_c\ri)$ in the phase plane ~ 
& ~ Parametric domain of physical relevance \\
\colrule
$J_1$: & $\le(\dfrac{\sq{3} - \sq{2 \fw + 3}}{\sq{2} ~ \fw}, \, 0\ri)$ 
& $\fw \in \!\le(\!\! - \rfraa 3 2 , \infty\ri)$ \\ 
\colrule
$J_2$: & $\le(\dfrac{\sq{3} + \sq{2 \fw + 3}}{\sq{2} ~ \fw}, \, 0\ri)$ 
& $\fw \in \!\le(\!\! - \rfraa 3 2 , \infty\ri) \cap \le\{0\ri\}$ \\
\colrule
$J_3$: & $\le(\dfrac 1 {\sq{6} \le(\fw + 1\ri)}, \, 0\ri)$ 
& $\fw \in \!\le[\! - \rfraa 4 3 , - \rfraa 6 5\ri]$ \\
\colrule
$J_4$: & $\le(\dfrac 1 {\sq{6} \le(\fw + 1\ri)}, \, 
\pm \dfrac{\sq{(2 \fw + 3) (3 \fw + 4)}} {\sq{6} \le(\fw + 1\ri)}\ri)$ 
& $\fw \in \!\le[\! - \rfraa 4 3 , \infty\ri) \cap \le\{- 1\ri\}$ \\
\colrule
$J_5$: & $\le( - \sq{\dfrac 3 2}, \, \pm \sq{\dfrac{3 \fw + 4} 2}\ri)$ 
& $\fw \in \!\le\{\!\! - \rfraa 4 3\ri\}$ \\ 
\botrule
\end{tabular} 
\la{J-tab-roots} }
\end{table}
\begin{table}[phtb]
\tbl{Eigenvalues of the linear perturbation matrix $\cM$ at the critical 
points, and the type and nature of these points in the Jordan frame.}{
\centering
\renewcommand{\arraystretch}{1.25}
\begin{tabular}{@{}cccc@{}} \toprule
CP & Eigenvalues of $\cM$ at CP & For parametric range: & CP type (nature) \\
\colrule
$J_1$ & $\m_1 = \m_2 = \dfrac 3 \fw \!\le[1 + \fw - \sq{1 + \dfrac{2 \fw} 3}\ri]$ 
& $\fw \in \!\le(\!\! - \rfraa 3 2 , \infty\ri)$: & Nodal Source (Unstable) \\ 
\colrule
\multirow{3}{*}{$J_2$} 
& \multirow{3}{*}{$\m_1 = \m_2 = \dfrac 3 \fw \!\le[1 + \fw + \sq{1 + \dfrac{2 \fw} 3}\ri]$} 
&  $\fw \in \!\le(\!\! - \rfraa 3 2 , - \rfraa 4 3\ri) \cup \le(0, \infty\ri)$: 
& Nodal Source (Unstable) \\
&  & $\fw \in \!\le\{\!\! - \rfraa 4 3\ri\}$: & Indeterministic \\
&  & $\fw \in \!\le(\!\! - \rfraa 4 3 , 0\ri)$: & Nodal Sink (Stable) \\
\colrule
\multirow{2}{*}{$J_3$} 
& \multirow{2}{*}{$\m_1 = - \m_2 = \dfrac{3 \fw + 4}{2 \le(\fw + 1\ri)}$} 
& $\fw \in \!\le\{\!\! - \rfraa 4 3\ri\}$: & Indeterministic \\
& & $\fw \in \!\le(\! - \rfraa 4 3 , - \rfraa 6 5\ri]$: & Saddle point (Unstable) \\
\colrule
\multirow{3}{*}{$J_4$} & \multirow{3}{*}{$\m_1 = \m_2 = -\, \dfrac{3 \fw + 4}{\fw + 1}$} 
& $\fw \in \!\le\{\!\! - \rfraa 4 3\ri\}$: & Indeterministic \\
& & $\fw \in \!\le(\! - \rfraa 4 3 , - 1\ri)$: & Nodal Source (Unstable) \\
& & $\fw \in \!\le(- 1 , \infty\ri)$: & Nodal Sink (Stable) \\
\colrule
$J_5$
& $\m_1 = - \m_2 = \dfrac{\sq{3} \le(3 \fw + 4\ri)}{\sq{2 \fw + 3}}$ 
& $\fw \in \!\le\{\!\! - \rfraa 4 3\ri\}$: & Indeterministic \\
\botrule
\end{tabular}
\la{J-tab-evalues} }
\end{table}
Table \ref{J-tab-evalues} shows the eigenvalues ($\m_1, \m_2$) of the linear perturbation 
matrix $\cM$ at each CP, and the type and nature of the CPs they determine (see subsection 
\ref{sec:E-aut}). 

Similar to what have seen in the Einstein frame, at most two of the five CPs (viz. $J_2$ and 
$J_4$ here) could be stable points, whereas one CP (viz. $J_3$) could be a saddle point. 
Particularly intriguing are the CPs $J_3$ and $J_4$, just as their Einstein frame counterparts 
$E_3$ and $E_4$ respectively, in pointing out the difference of the MST-cosmological dynamics 
over that of the non-interacting quintessence and dust models \cite{cope,tsuj}. Whereas the 
asymptotic forms of the solutions represented by $J_3$ imply co-existence of the dust and the 
field $\t$, those represented by $J_4$ imply cosmic acceleration\footnote{Verify using Eq. 
(\ref{J-w}) that at the CP $J_4$, the total EoS parameter of the system is $\wJ = - 1$, i.e. 
the acceleration condition $\wJ < - \rfraa 1 3$ is always satisfied.} resulting from a 
$\t$-dominance (the equivalent of the Einstein frame DE-dominance). Also the evolution of the 
universe leading up to $J_4$ has qualitative similarity with that leading up to the stable 
point $E_4$ in the Einstein frame.

There are some differences though in the interpretation of the results of the Jordan 
frame and Einstein frame analyses. Note in particular that for $\fw \in (-1, 0)$, the 
CPs $J_2$ and $J_4$ are both stable points of physical relevance, whereas in the Einstein 
frame we never had two stable points in any given parametric domain. Moreover, unlike its 
Einstein frame analogue $E_2$, the stable point $J_2$ supports solutions exhibiting cosmic 
acceleration\footnote{Verify using Eq. (\ref{J-w}) that at the CP $J_2$, $\wJ = 1 + 
\le(\rfraa 2 \fw\ri) \le[1 + \sq{1 + \rfraa{2 \fw} 3}\ri]$, so that in the domain $\fw \in 
(-1, 0)$ the acceleration condition $\wJ < - \rfraa 1 3$ is satisfied.} at the asymptotic 
limit (as do the stable point $J_4$). So for $- 1 < \fw < 0$ (whence there are two disjoint 
hyperbolic regions $\fM$) we require to determine first which among $J_2$ and $J_4$ is the 
appropriate stable CP at which the physical trajectories would terminate. Whereas at $J_4$ 
the total EoS parameter of the system $\wJ = - 1$ always, it is easy to check that at 
$J_2$ one has $\wJ < -1$ for $- 1 < \fw < 0$. Therefore, in this domain the trajectories 
terminating at $J_2$ would imply that in the asymptotic limit the universe is at a {\em 
super-accelerating} or {\em phantom} state which is so strong that even $\wJ < - 1$. A 
super-acceleration with a component EoS parameter, such as that for $\t$, less than $- 1$ 
is acceptable. But if the system's EoS parameter $\wJ < - 1$, it may lead to physical 
instabilities (against the cosmological metric perturbations \cite{mukh,brand}). So there 
is a strong reason to discard $J_2$. Besides, $J_2$ being always shifted from the origin 
(along the $X$-axis), the trajectories which represent a nearly $\L$CDM evolution cannot 
terminate at $J_2$. The reason is that the $\L$CDM configuration (recovered in the limit 
$\fw \rightarrow \infty$) is represented by a line segment of finite size along the 
$Y$-axis and symmetric about the origin. 

In fact, within the physically admissible zone $\fM$, any point on the $Y$-axis implicates 
a configuration of a non-dynamical $\t$, whose potential (originating from the torsion 
pseudo-trace $\cA^\m$) plays the role of a cosmological constant $\L$. In other words, 
for any point on the $Y$-axis, $\cA^\m$ is the key torsion constituent whose contribution 
to the critical density of the universe has a fixed value, reminiscent of that due to 
$\L$. On the other hand, the points which are not located on the $Y$-axis represent the 
system configurations for a dynamical $\t$. The extent of such dynamics is determined 
by the magnitude of $X$, or the torsion trace mode $\cT_\m$, because in the Jordan frame 
we have the torsion trace parameter $\le|\cT\ri| = \dot{\t} \sim H X$. 

As to solving numerically the autonomous Eqs. (\ref{J-auteq1})--(\ref{J-auteq2}), since 
we have the admissible region $\fM$ bounded from both inside and outside now, the choice 
of initial conditions for the phase plane trajectories are much restricted compared to 
that in the Einstein frame. Also, for simplicity in the numerical calculations and to 
emphasize on the viable MST-cosmologies not much deviated from $\L$CDM, we resort to the 
situations corresponding to $\fw > 0$ only, whence $\fM$ has closed (elliptic or circular) 
boundaries. Figs. \ref{J-PPfig} (a)--(e) show the evolutions of some select physical 
trajectories in the $XY$ phase plane, for the exemplary cases $\fw = 0.5, 1, 5, 25$ and 
$50$. 
\begin{figure}[h]
\centering
\subfloat[{\footnotesize $\fw = 0.5$}]{\includegraphics[scale=0.51]{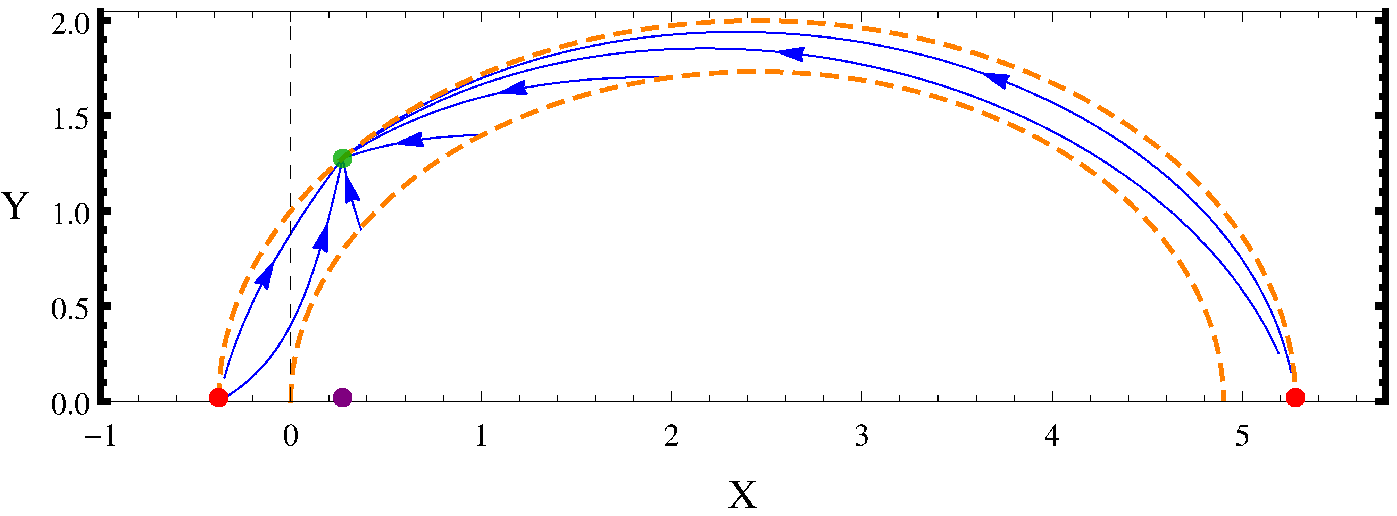}}\quad
\subfloat[{\footnotesize $\fw = 1$}]{\includegraphics[scale=0.485]{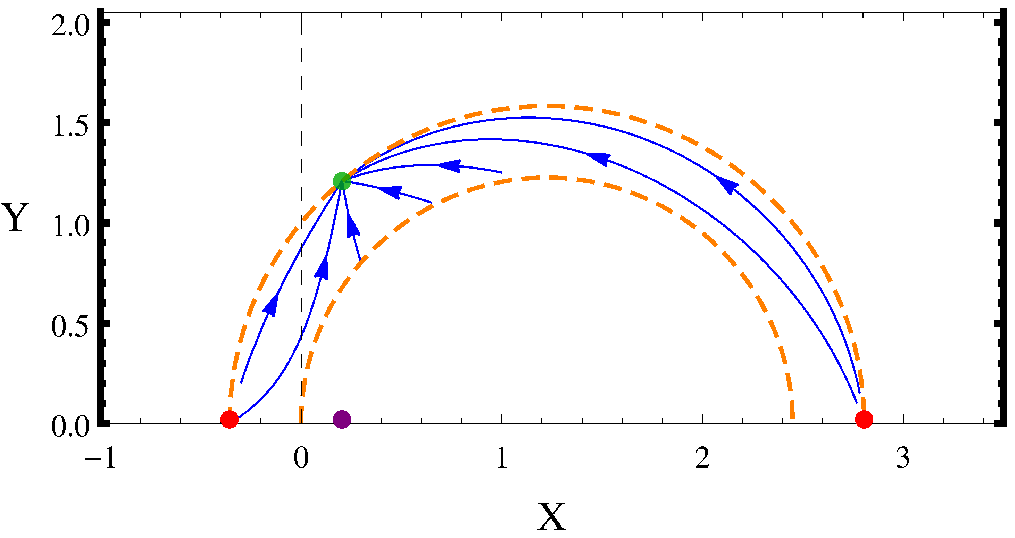}}\\
\subfloat[{\footnotesize $\fw = 5$}]{\includegraphics[scale=0.4965]{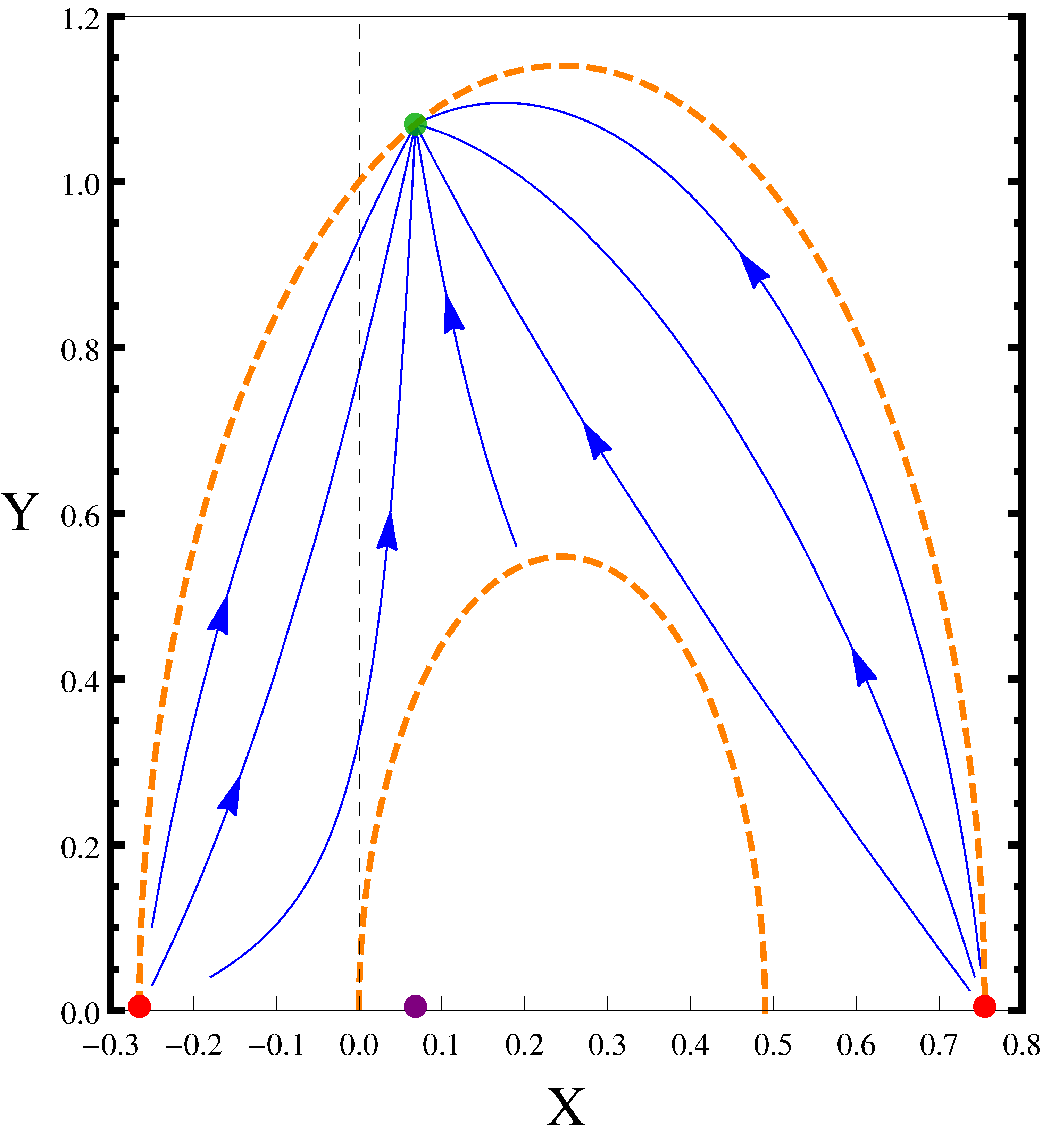}}\quad
\subfloat[{\footnotesize $\fw = 25$}]{\includegraphics[scale=0.51]{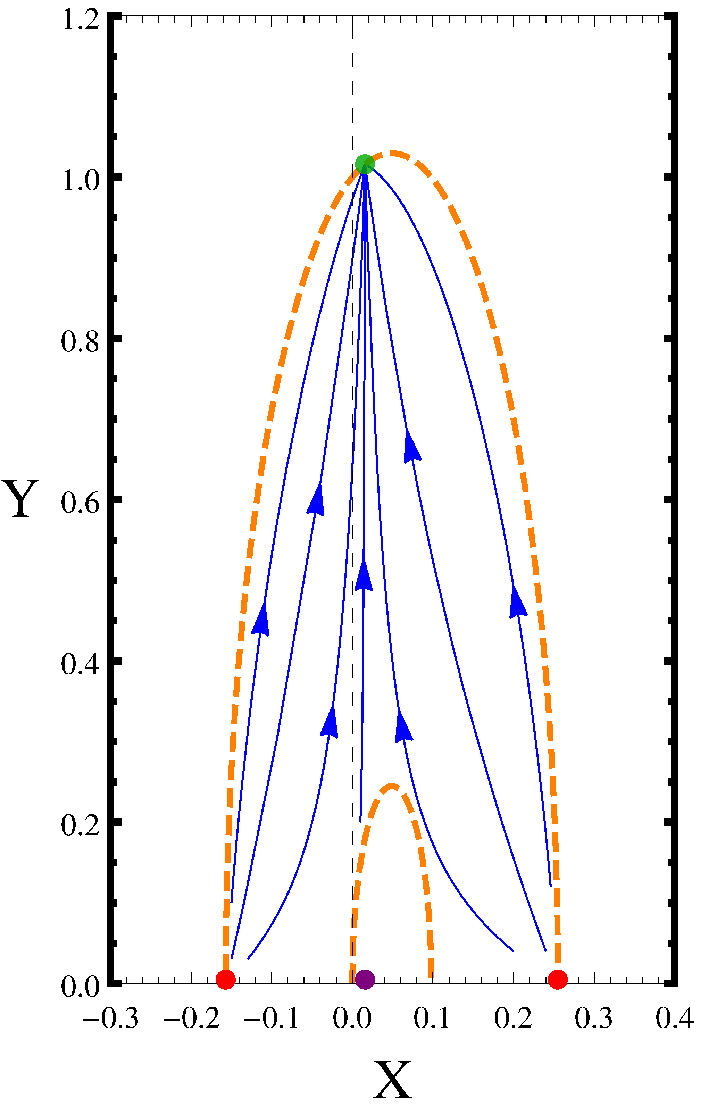}}\quad
\subfloat[{\footnotesize $\fw = 50$}]{\includegraphics[scale=0.5]{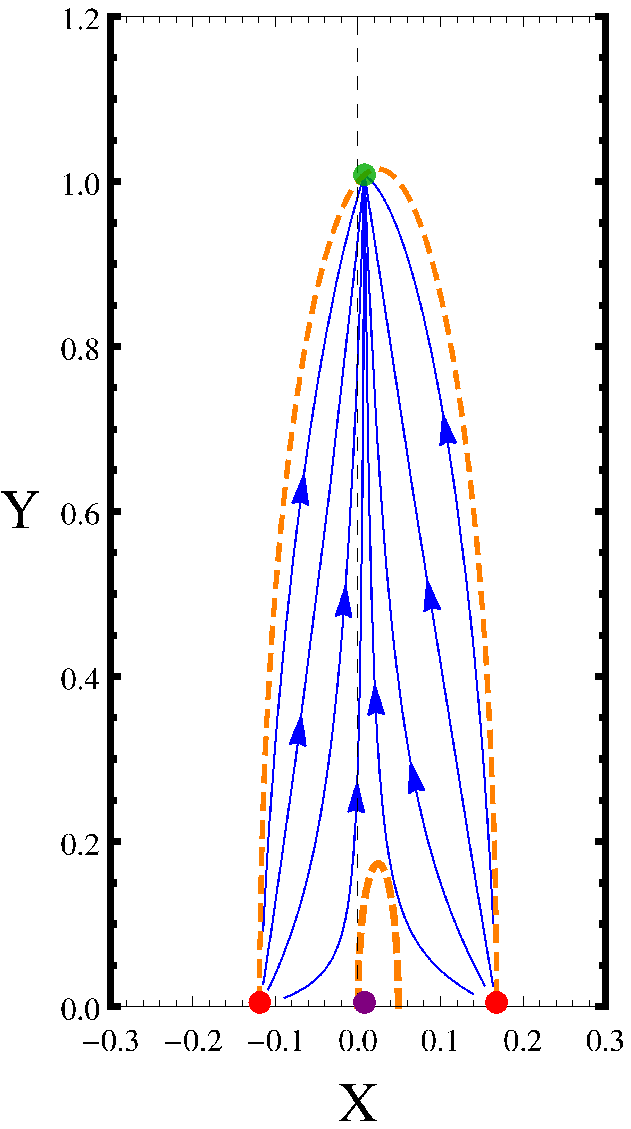}}\quad
\caption{\footnotesize Jordan frame phase portraits for different values of the parameter 
$\fw$, viz. $0.5, 1, 5, 25$ and $50$. The dots represent the critical points, arrows mark 
the direction of time-evolution of trajectories and the dashed curves demarcate the region 
of phase plane which supports cosmologies with a non-negative (conventional) matter density. 
As $\fw$ increases, the saddle point and the stable point shift to the left along the abscissa 
and the circumference of the outer boundary respectively.} 
\la{J-PPfig}
\end{figure}
We see that the phase portraits are quite similar to those in the Einstein frame (see Figs.
\ref{E-PPfig} (a)--(e)), except of course the presence of the inner boundary of $\fM$, and 
that the size of both the boundaries reduce with increasing $\fw$ (i.e. decreasing $\b$) not 
only along $X$ but also along $Y$. The trajectories terminate at the stable point $J_4$ on 
the outer boundary. On the $X$-axis, and vertically beneath $J_4$, we have the saddle point 
$J_3$ which, although remains outside the admissible region $\fM$, has significance in 
funnelling the trajectories towards $J_4$. As $\fw$ increases, both $J_3$
and $J_4$ move towards the origin (with the shrinkage of $\fM$), and so do the unstable
points $J_1$ and $J_2$ at the two intersections of the outer elliptic boundary and the 
$X$-axis. The other CP $J_5$ (not shown in Figs. \ref{J-PPfig} (a)--(e)) is beyond the 
outer boundary of $\fM$ for all values of $\fw > 0$.  

\subsection{Dynamical evolution in the effective scenario} \la{sec:J-eff}

The task that remains now is to verify whether the dynamical evolution of the universe at 
a stable point is in accord with the exact solution we have found in the paper I by 
explicitly solving the Jordan frame MST-cosmological equations \cite{ssasb2}. To do 
this, it is imperative to consider the effective scenario in the Jordan frame, in which 
we have critical density $\r$ [{\it cf}. Eq. (\ref{J-criteff})] that satisfies the usual 
conservation relation (\ref{J-consveff}). In fact, so far all our arguments have been 
based on the Jordan frame matter density parameter $\Om_J$ or(and) the total EoS parameter 
$\wJ$. The interpretations of the dynamical evolution of the universe in terms of an 
effective DE constituent is facilitated by the decomposition (\ref{J-criteff}) of the 
(effective) critical density $\r$ in the dust-like matter density $\rmt$ and the DE 
density $\rx$. Let us define the corresponding density parameters respectively as $\Om := 
\rfraa{\rmt\!\!}{\r}$ and $\OX := \rfraa{\rx}{\r}$, and also the EoS parameters for the 
DE and the system respectively as $\wx := \rfraa{\px}{\rx}$ and $\sw := \rfraa{p}{\r}$.
Then from Eqs. (\ref{J-criteff}), (\ref{J-de-denspr}), (\ref{J-m-dens}) and (\ref{J-w}) 
we have
\bea 
&& \OX = 1 - \Om \,\,, \qquad \wx = \fr{\sw}{\OX} \,\,; \quad \mbox{with} \quad \sw = \wJ 
\,\,, \qquad \mbox{and}
\la{J-eff-DE} \\
&& \Om =\, \Om_J \, e^{\sq{6} \, F} \,= \le(1 \,-\, \fw \, X^2 \,-\, Y^2 
\,+\, \sq{6} \, X\ri) e^{\sq{6} \, F} \,\,, 
\la{J-mf-dens} 
\eea
where
\be \la{J-tauF}
F (N) \,\equiv\, \int_0^N X(\cN) \, d\cN  \,\,, \qquad \mbox{implying} \qquad
\t (N) \,=\, \fr{3 \sq{6}} 2 \, F (N) \,\,.
\ee
What the exponential factor $e^{\sq{6} \, F}$ in Eq. (\ref{J-mf-dens}) does is that it brings 
back the situation we have had in the Einstein frame (see subsection \ref{sec:E-aut}). That is,
the matter density parameter $\Om$ would keep on evolving with time (or $N$) even after the 
system reaches a CP. So the condition $\Om < 1$ would get violated eventually (at some later 
epoch), unless we set $\Om = 0$ in the asymptotic limit. Hence, the physical relevance of the 
phase plane trajectories requires the corresponding CPs to lie on the curve whose equation is
\be \la{J-cp-exist}
\fw \, X_c ^2 \,+\, Y_c ^2 \,-\, \sq{6} \, X_c \,=\, 1 \,\,.
\ee
This curve therefore encloses (or demarcates) the physically admissible region(s) $\fM$ in 
the phase. Note also that this curve is actually the outer boundary (or demarcation line(s)) 
of $\fM$ we have had while working in terms of the conventional matter density parameter 
$\Om_J$ (in subsections and \ref{sec:J-aut} and \ref{sec:J-dyn}). We do not have any inner 
boundary of $\fM$ now, or we may say that the inner boundary is {\em trivial} (i.e. coincident 
with the origin $(0,0)$) as in the Einstein frame. Nevertheless, the situation here still has 
one difference with that in the Einstein frame. That is, with increasing (presumably positive) 
values of $\fw$ (i.e. with decreasing $\b$), the size of $\fM$ shrinks along both $X$ and $Y$ 
(instead of along $X$ only, as in the Einstein frame). As to the phase portraits in Figs. 
\ref{J-PPfig} (a)--(e), since $\fM$ gets enlarged with the removal of the inner boundary, 
we have more choices for obtaining the trajectories numerically. Accordingly the saddle point 
$J_3$ (situated on the $X$-axis) now has more trajectories to deflect towards the stable point 
$J_4$ at the boundary (\ref{J-cp-exist}).

Now, the cosmic acceleration being the main aspect of the exact Jordan frame solution found 
in paper I, one expects this solution to transpire to one of the stable 
points $J_2$ and $J_4$ in the asymptotic limit, since they are the only CPs that support 
accelerating cosmologies. By inspection we find that $J_4$ is the appropriate one, since at 
this CP $\le(X_c = \rfraa n {\!\!\!\sq{6}}, \, Y_c = \sq{\le(1 + \rfraa n 2\ri) \le(1 + 
\rfraa n 3\ri)}\ri)$, where $n = (1 + \fw)^{-1}$, the torsion scalar is given by $e^\t = 
a^{\rfra{3 n \!}{\! 2}}$, whence the Hubble parameter:
\be \la{Hub-J4}
H^2 \,=\, \fr{2 \k^2 \, \L}{\le(n + 2\ri) \le(n + 3\ri)} \,\,, 
\ee
turns out to be precisely the asymptotic (i.e. the $a \rightarrow \infty$ limiting) form of 
that we have had in paper I, while deriving the exact solution in the 
Jordan frame (see section 5.1 therein). The stability of such a solution is thus established.

Let us finally look into the evolution of say, the EoS parameters $\wx (N)$ and $\sw (N)$ 
over some fiducial trajectory, typically for two cases $\fw = 5$ and $\fw = 50$ (see Figs.
\ref{J-wwx} (a) and (b) respectively). As before, we take this trajectory (in either of 
these cases) to be the one in compliance with the initial conditions $X (0)$ and $Y (0)$, 
set with $\Omp = 0.3$ and keeping in mind the power-law ansatz $\, e^\t = a^{\rfra{3 n \!}
{\! 2}}$ for the stable solution in \cite{ssasb2}. Eqs.(\ref{J-aut}) and (\ref{J-m-dens}) 
then give
\be \la{J-ini}
X (0) \,=\, \fr n {\sq{6}} \,\, \qquad \mbox{and} \qquad 
Y (0) \,=\, \sq{\le(1 + \fr n 2\ri) \le(1 + \fr n 3\ri) - \Omp} \,\,.
\ee
With $n = (\fw + 1)^{-1}$, we therefore have $\le(X (0), Y (0)\ri) = (0.068, 0.918)$ for 
$\fw = 5$, and $\le(X (0), Y (0)\ri) = (0.008, 0.846)$ for $\fw = 50$.
\begin{figure}[h]
\subfloat[{\footnotesize $\fw = 5$}]{
{\def\big{\includegraphics[width=5.7cm,keepaspectratio=true]{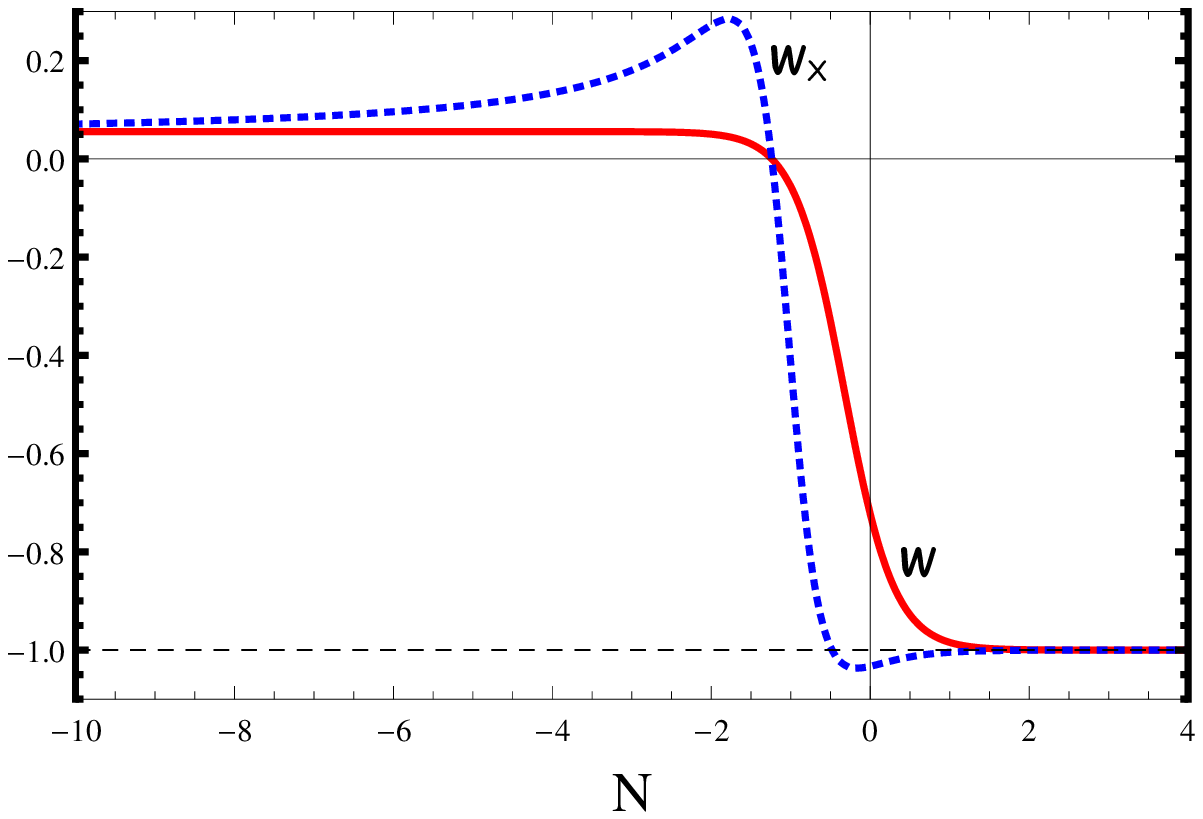}}
   \def\little{\includegraphics[height=2.3cm,width=3cm]{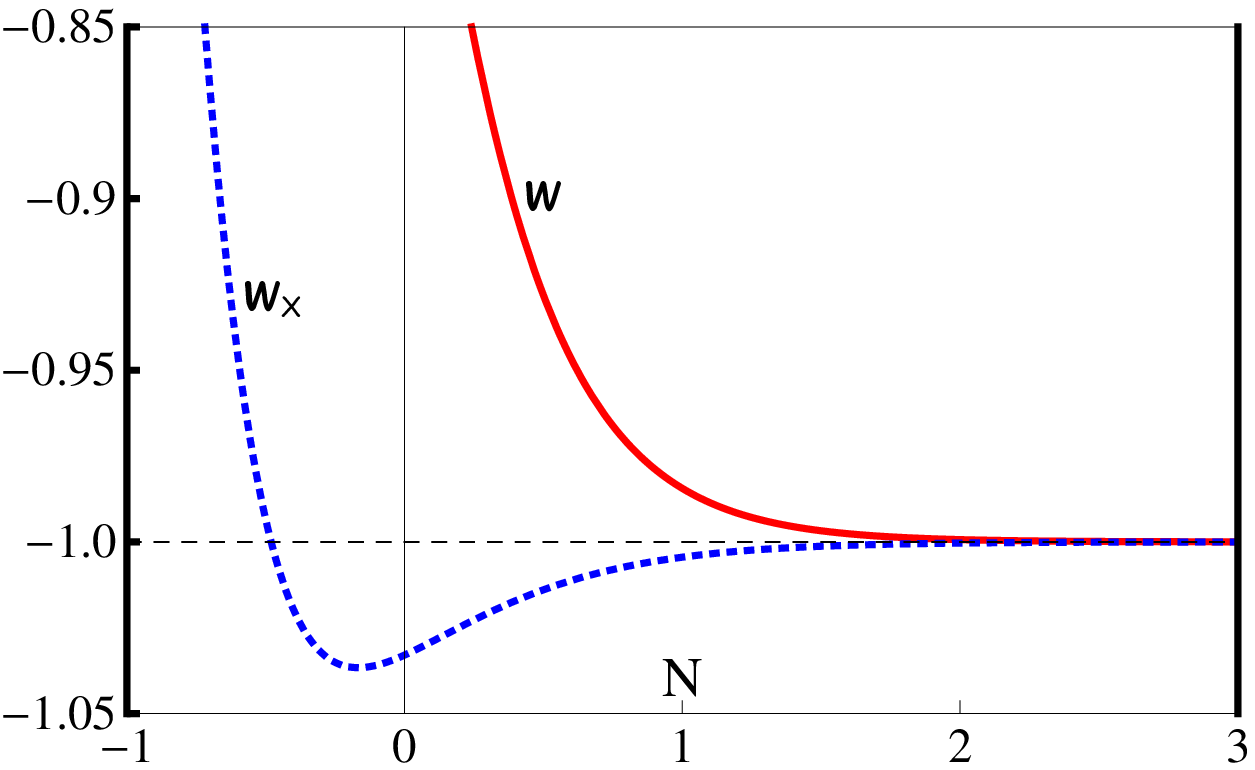}}
   \def\stackalignment{l}
   \topinset{\little}{\big}{22pt}{11pt}}}\quad  
\subfloat[{\footnotesize $\fw = 50$}]{
{\def\big{\includegraphics[width=5.7cm,keepaspectratio=true]{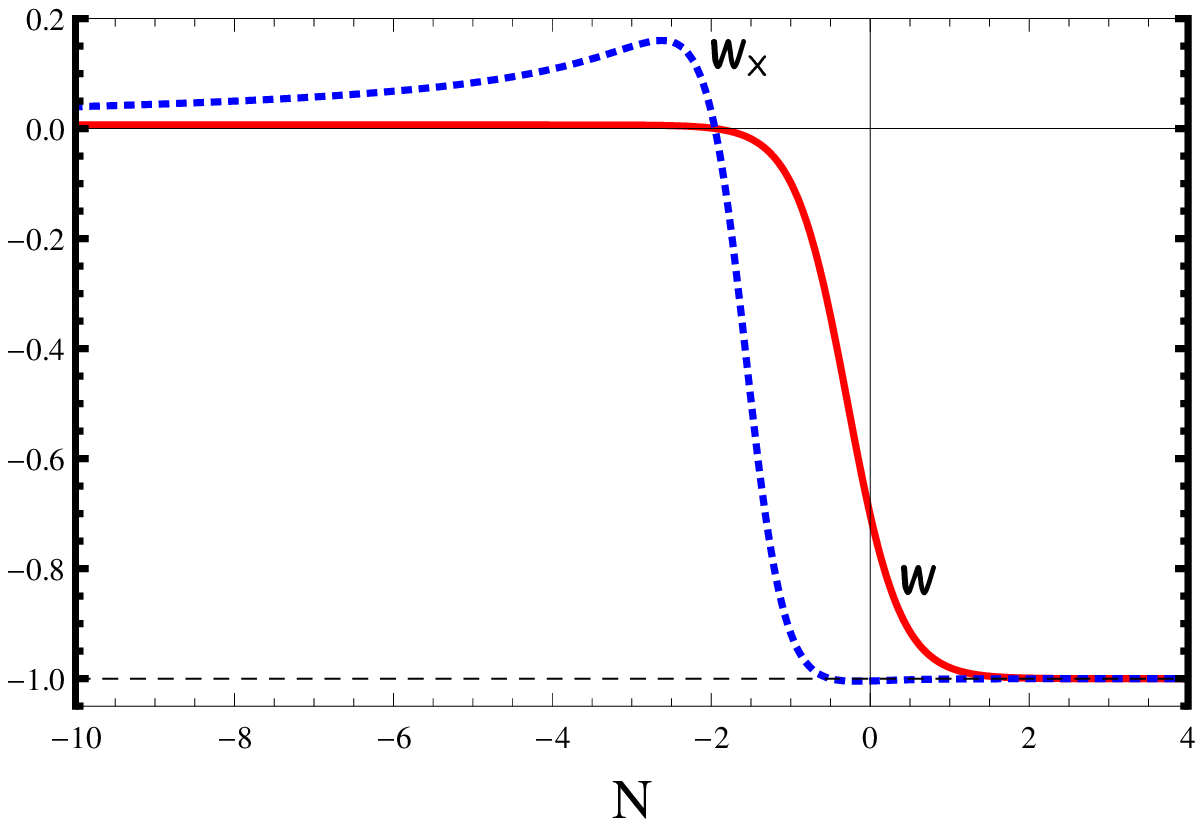}}
   \def\little{\includegraphics[height=2.35cm,width=3cm]{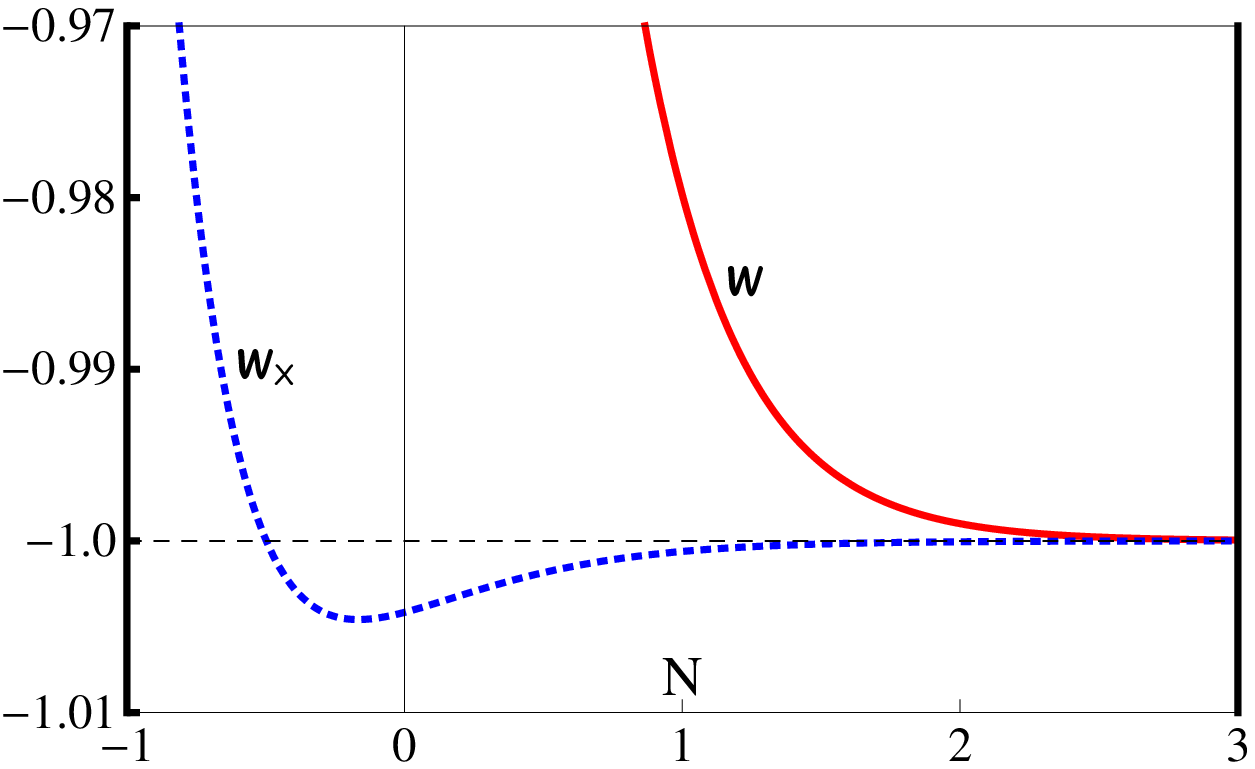}}
   \def\stackalignment{l}
   \topinset{\little}{\big}{24pt}{11pt}}}
\caption{\footnotesize Evolution of the EoS parameters $\wx$ and $\sw$, for the DE and the system 
respectively, over a fiducial trajectory with initial conditions: (a) $X_i \equiv 0.068$ and $Y_i 
\equiv 0.918$ for $\fw = 5$, and (b) $X_i \equiv 0.008$ and $Y_i \equiv 0.846$ for $\fw = 50$. 
Portions of the plots, exaggerated in the range $N \in (-1, 3)$ are shown in the insets of (a) 
and (b).}
\la{J-wwx}
\end{figure}
The EoS parameter $\sw$ for the system evolves in a similar way as in the Einstein frame, viz. 
remains almost fixed at a small positive value deep in the past, and then in the near past 
makes a steep transition to a value very close to $-1$, at which it saturates. At the present 
epoch ($N = 0$) the universe is in the transition phase, which ends in the near future. The 
smaller the value of the parameter $\fw$, the steeper is the transition. The EoS parameter $\wx$ 
for the DE, on the other hand, also shows a transition from a positive value to a negative value, 
which is even more steep (compared to that of $\sw$ for the same value of $\fw$). Deep in the 
past $\wx$ increases very slowly with $N$ from a small positive value, until reaching a maximum 
in the near past. Then it decreases rapidly and attains a minimum value {\em less than} $-1$,
increases again and asymptotically tends to $-1$ (from below). The transition to the minimum 
value ends before the present epoch, whence the universe continues to be in the {\em 
super-accelerating} or {\em phantom} regime (characterized by $\wx < -1$)\footnote{Note however
that the total EoS parameter $\sw$ of the system never goes below $-1$. Therefore, physical 
instabilities against metric perturbations \cite{mukh,brand} can not arise.}. Such an effective 
super-acceleration is actually the striking feature of the Jordan frame solution obtained in
paper I. Furthermore, the smaller the value of $\fw$ the steeper is the transition of $\wx$ 
from its maximum value to its minimum value, and the greater are these values in magnitude (see 
the exaggerated portions shown in the insets of Figs. \ref{J-wwx} (a) and (b)).

\section{Conclusion} \la{sec:concl}

We have thus made a systematic analysis of the dynamical stability of cosmological DE solutions 
in the scalar-tensor equivalent formulation of a non-minimal metric-scalar coupling with torsion 
(viz. the MST coupling). The roles of the individual torsion modes, viz. the trace $\cT_\m$ and 
the pseudo-trace $\cA_\m$, on the cosmological dynamics are envisaged by working in terms of a 
suitably defined {\em torsion scalar} $\t$ and a {\em torsion constant} $\L$. Apart from $\t$, 
we have taken for simplicity just a single component cosmological matter, viz. the pressureless 
`dust', thus restricting our analysis to a two-dimensional phase space, or the {\em phase plane}. 
Not only the analysis applies to the exact solutions found in paper I, in Einstein and Jordan 
frames \cite{ssasb2}, but also to other solutions plausible for a cosmic acceleration in the 
MST-cosmological setup. Essentially, we have had the Einstein frame and Jordan frame cosmological
equations cast in the form of the respective autonomous sets, whose equilibrium solutions (in the
asymptotic limit) are the {\em critical points} (CPs). Identifying the type and nature of these 
CPs, by applying the linear perturbation theory, we have analysed the phase plane trajectories of 
physical relevance, and the solutions they support, satisfying the requirement of cosmic acceleration 
in the asymptotic limit. 

In the Einstein frame, we have found as many as {\it five} CPs existing, including stable and 
saddle points. Nevertheless, the location of these CPs in the phase plane, and the cosmological
dynamics leading up to them, are quite different from that in the case of a quintessence field 
(with an exponential potential) in presence of dust \cite{cope,tsuj}. The reason is two-fold: 
firstly, we have the torsion scalar $\t$ interacting with the cosmological matter in the Einstein 
frame MST setup, and secondly, the phase variables need to be such that their calibrations are 
not affected by the MST coupling parameter $\b$. For convenience, we have chosen to work in an 
effectively non-interacting scenario, in which the physical acceptability of cosmological 
solutions crucially depends on the eventual extinction of the (dust-like) matter. This implies 
that the CPs are of physical relevance only when they lie on an ellipse centred at the origin. 
Such an ellipse therefore forms the boundary of the physically admissible region $\fM$ for the 
phase plane trajectories, at a given value of $\b$. With increasing $\b$, the eccentricity of the 
ellipse decreases, and the area of $\fM$ increases, whence the CPs move further and further away 
from the origin. Of prime importance is the domain $0 < \b < \rfra 1 2$, within which four CPs 
exist, including the stable point that supports solutions exhibiting cosmic acceleration 
asymptotically. An exact DE solution of such sort is that found in paper I. We have not only 
verified the stability of this solution, but also have numerically obtained the evolution profiles 
of the DE density and EoS parameters, $\OX$ and $\wx$, over a fiducial trajectory whose initial 
conditions are set in accord with such a solution. Nevertheless, our emphasis has mostly been on 
a much smaller value of $\b$ ($\sim 10^{-1}$ or lesser), so that the DE evolution remains within 
the proximity of the concordance $\L$CDM model. 

In the Jordan frame, we have had the option to resort to either the conventional scenario in 
which the critical density of the universe is not conserved and depends explicitly on the running
gravitational coupling factor, or the effective scenario in which the critical density is by
definition conserved and decomposed into the densities due to the dust and a DE component, which 
are individually conserved. Although the effective scenario is more useful from the observational
perspective, we have carried out the dynamical analysis first in the conventional scenario, which 
is in line with the standard scalar-tensor approaches in the literature \cite{tsuj}. In due 
course, we have shown that the general results of such an analysis do not differ in the effective
scenario, except the broadening of the admissible regions $\fM$ for physical trajectories in the 
phase plane (see the full account given in \S \ref{sec:J-aut} on the boundaries of such regions 
at different domains of the effective Brans-Dicke (BD) parameter $\fw \, (\sim \b^{-1})$). We have 
found five distinct CPs existent, as in the Einstein frame. Examining the type and nature of these 
CPs and the solutions they support, we have once again identified the stable point at which the 
universe remains in an accelerating state of expansion asymptotically, without any physical 
instability against cosmological metric perturbations. In the effective scenario, we have not 
only verified that the exact Jordan frame solution found in paper I indeed transpires to this 
stable CP in the asymptotic limit, but also have worked out numerically the evolution profiles of 
the EoS parameters $\sw$ and $\wx$, for the system and the DE respectively, over fiducial 
trajectories leading up to the stable CP. While the evolution of $\sw$ is found to be similar 
to that in the Einstein frame, $\wx$ has shown the striking feature of crossing the {\em phantom 
barrier} at an epoch in the recent past.

On the whole, the main aspect of the MST-cosmological analysis is the indispensable role of the 
torsion pseudo-trace mode $\cA^\m$ in the stable DE configurations, in both Einstein and Jordan 
frames. The trace mode $\cT_\m$ (or its scalar field source $\f$), although crucial in the MST
formalism, only has a supplementary role in the DE evolution. This is evident from the fact that 
the effective measure of $\cA^\m$, viz. the torsion constant $\L$, ensures the culmination of a 
phase trajectory at the stable point that supports the accelerated expansion of the universe.
On the other hand, the torsion scalar $\t$, derived from $\cT_\m$, is responsible for a dynamical
evolution of the DE. The dynamics is weak enough as long as the coupling parameter $\b$ is small.
While in the Einstein frame we have smallness of $\b$ acquiescing to the stringent bound $\b < 
\rfra 1 2$, in the Jordan frame it implies a large BD parameter $\fw$, which is in accord with 
many independent studies \cite{BDbounds1,BDbounds2}. So it is a generic outcome of our analysis 
that stable DE models (including those found in paper I) \cite{ssasb2} result from a feeble 
metric-scalar coupling with torsion (no matter how weak the latter is).

In principle, the dynamical analysis carried out here can be extended to that applicable to any 
scalar field DE model in which the scalar interacts with the dust-like matter. The effective 
scenario is best suited for this. One has to simply consider the (usual) critical density, viz. 
$\r := \rfraa{3 H^2 \!\!}{\!\! \k^2}$, and its splitting into two non-interacting components --- 
the dust and whatever is left over (let that be the DE), and carry out the analysis just as in 
this work. However, note that the results of the analysis would always be model specific, because 
the phase variables for any given model are required to be so chosen that they are {\it not} 
explicitly dependent on a system parameter, different domains of which have diverse implications 
in the analysis. In that sense, the results obtained in this paper have their {\it uniqueness} 
attributed to the analysis we have carried out by keeping the phase variables independent of our 
system parameter $\b$ (or $\fw$) explicitly. Of actual importance is of course our definition of 
the phase variables in terms of the torsion scalar $\t$ or the torsion constant $\L$, rather than 
in terms of a redefined quintessence-like scalar field or its potential. 

Let us end this paper with some open issues, such as: (i) what happens if instead of the torsion
constant $\L$, we consider some other contribution of the torsion pseudo-trace $\cA^\m$ (sourced
by say, the Kalb-Ramond axion in string-inspired scenarios \cite{pmssg,saa})? (ii) what if we
introduce an interaction of a generic form between torsion and the cosmological dust (or any other 
perfect fluid matter)? (iii) how about correlating the results of the dynamical analysis here for 
our MST-cosmological system, with that for the models of coupled quintessence, coupled tachyon 
etc. (as for e.g. in some recent works \cite{land} and references therein)? (iv) how about 
extending the dynamical analysis in this paper to say, the Chaplygin gas or Chameleon cosmologies 
in the MST framework, or to generalize it for the modified gravity theories, such as $f (\cR)$ or 
mimetic gravity \cite{fR,modgrav,mimgrav}, in presence of torsion? and so on. Works addressing some 
of these are in progress \cite{ssasbCham,ssetalDDE} and we hope to report them soon.

\section*{Acknowledgements}

The work of ASB was supported by the Council of Scientific and Industrial Research (CSIR), Government of India. SS acknowledges the R \& D Grant DRCH/R \& D/2013-14/4155, Research Council, University of Delhi.

\end{document}